\documentclass[lettersize,journal]{IEEEtran}
\usepackage{cite}
\usepackage{amsmath,amssymb,amsfonts}
\usepackage{algorithmic}
\usepackage{graphicx}
\usepackage{textcomp}
\usepackage{wrapfig}
\usepackage[T1]{fontenc}
\usepackage[utf8]{inputenc}
\usepackage[english]{babel}
\usepackage{xcolor}

\def\BibTeX{{\rm B\kern-.05em{\sc i\kern-.025em b}\kern-.08em
    T\kern-.1667em\lower.7ex\hbox{E}\kern-.125emX}}

\begin{document}
\title{Secrecy Performance of an Artificial Noise Aided Dual-Hop SWIPT MIMO-NOMA Network in Presence of Practical Impairments}
\author{Büşra Demirkol, Oğuz Kucur, and Güneş Karabulut Kurt, \IEEEmembership{Senior Member, IEEE}
\thanks{The authors thank TÜBİTAK for supporting this work with the grant number 121E467. (Corresponding author: Büşra Demirkol.) }
\thanks{Büşra Demirkol, Oğuz Kucur, Department of Electronics Engineering, Gebze Technical University, Kocaeli, Türkiye (e-mail: busrademirkol@gtu.edu.tr; okucur@gtu.edu.tr).
Güneş Karabulut Kurt, Department of Electrical Engineering, Polytechnique Montréal, Montréal, Canada (e-mail: gunes.kurt@polymtl.ca).}
}

\maketitle
\begin{abstract}
This paper investigates the secrecy performance of a cooperative non-orthogonal multiple access (NOMA) network under practical impairments. In the considered system, the source employs orthogonal space-time block coding (OSTBC) transmission, legitimate users adopt receive antenna selection (RAS), and the eavesdropper utilizes maximal-ratio combining (MRC) scheme. Moreover, an energy-constrained amplify-and-forward relay harvests energy from the received signal and allocates the harvested energy to information forwarding and artificial noise (AN) generation. To capture the realistic performance of a more practical transmission scenario, channel estimation errors (CEE), feedback delay (FBD), imperfect successive interference cancellation (SIC), and imperfect AN cancellation are taken into account. 
For this system, closed-form expressions for the legitimate-user signal-to-interference+noise ratios (SINR) cumulative distribution function (CDF) and the eavesdropper SINR probability density function (PDF) are derived, and a tractable analytical secrecy outage probability (SOP) expressions are obtained. \textcolor{black}{In addition, the high-SNR secrecy behavior is characterized under ideal channel state information (CSI) conditions and practical impairments, revealing the achievable secrecy diversity order and the secrecy outage floor, respectively. Moreover}, secrecy throughput and energy efficiency are analyzed based on the obtained SOP results. The numerical findings are in close agreement with the analytical results and further illustrate how the secrecy behavior of the proposed system is affected by key design parameters and practical impairments. The obtained results indicate that transmit and receive diversity and proper AN power allocation significantly improve the secrecy performance, whereas practical impairments lead to non-zero SOP floors in the high signal-to-noise ratio region.
\end{abstract}

\begin{IEEEkeywords}
Artificial noise (AN), energy harvesting (EH), non-orthogonal multiple access (NOMA), orthogonal space-time block coding (OSTBC) technique, receive antenna selection (RAS), secrecy outage probability (SOP).
\end{IEEEkeywords} 

\section{Introduction} \label{sec:introduction}
\IEEEPARstart{T}{he} Internet of Things (IoT) has become a key application area for future wireless networks, where a very large number of devices are expected to exchange information with different traffic patterns and service requirements \cite{Nguyen2022IoTSurvey,Ahmed2024SecureIoT}. In such dense environments, many devices are low-power and low-complexity terminals, and their access to the network must be supported under limited spectrum and energy resources \cite{Elgarhy2024IoTURLLCmMTC}. Accordingly, spectrum and energy efficient transmission has become a key design objective for sixth-generation (6G) networks, especially for IoT and machine-type communication scenarios involving low-power and energy-constrained devices~\cite{Nguyen2022IoTSurvey, Ntabeni2024MTCEnergyEfficiency}. The coexistence of massive connectivity demands and limited spectrum resources makes conventional orthogonal access strategies less efficient when the number of active devices increases. Non-orthogonal multiple access (NOMA) addresses this limitation by enabling multiple users to access the same radio resources, thus improving spectral efficiency and supporting massive connectivity\cite{Ding2017,Aldababsa2018}. 

In power-domain NOMA, users are served over the same radio resources by combining their messages in a single superimposed transmit signal. The power assigned to each message is determined according to the channel conditions of users, and successive interference cancellation (SIC) is then applied at the receiver side to separate the desired message from the superimposed signal. This mechanism improves spectrum utilization and enables simultaneous access for multiple users, which makes NOMA suitable for dense IoT and massive connectivity scenarios \cite{Ding2017,Liu2017,Mondal2025BackComSurvey}. However, the performance of NOMA depends on suitable power allocation and successful SIC operation. In practical systems, residual interference caused by imperfect SIC (ipSIC) may degrade the received signal quality, especially for users that need to cancel multiple signal components.

To improve the coverage and reliability of NOMA networks, cooperative and relay-assisted NOMA schemes have been widely investigated. In cooperative NOMA, users with favorable channel conditions may assist the transmission of users with weaker links, whereas in relay-assisted NOMA, a dedicated relay can be employed to support the source-to-user transmission \cite{Ding2015,Indraganti2026CNOMASurvey}. Such relaying structures are particularly useful in IoT scenarios, where distant or cell-edge devices may experience severe attenuation and may not reliably decode the directly transmitted NOMA signal. In this direction, amplify-and-forward (AF) relay-aided NOMA-based IoT networks have been studied to improve throughput and energy efficiency under practical SIC limitations \cite{Jee2023AFRelayIoT}. Since relay nodes or IoT devices may operate under limited energy resources, simultaneous wireless information and power transfer (SWIPT) has also been integrated into cooperative NOMA systems to support energy-constrained relaying \cite{Nasir2013}. In SWIPT-assisted NOMA relaying, the harvested energy can be used for forwarding the received information, which makes the system more suitable for energy-limited IoT and relay-assisted deployments.

Due to the broadcast nature of wireless propagation, confidential information transmitted over networks may be intercepted by unintended receivers. This issue becomes more critical in NOMA-based IoT networks since superimposed signals contain the messages of multiple users and SIC may allow a receiver to decode signals intended for other users. Therefore, physical-layer security (PLS) has emerged as a complementary approach to conventional cryptographic techniques by exploiting the randomness of wireless channels to protect confidential transmission \cite{Wyner1975,Bloch2011,Mukherjee2014}.
Since many IoT devices have limited computational capability and strict energy constraints, PLS is particularly attractive for IoT-oriented networks where lightweight security mechanisms are required. Secure SWIPT-NOMA transmission has been investigated under both direct and cooperative communication scenarios in the literature. In particular, single-hop secrecy networks considering practical impairments and diversity techniques are studied in \cite{Demirkol2026TVT,Ganning2017}.

The secrecy performance of NOMA and cooperative NOMA systems has been investigated in several recent studies. Relay selection schemes for cooperative NOMA networks are studied in \cite{Lei2019}, where secrecy outage probability (SOP) expressions are derived over Nakagami-\(m\) fading channels. For cooperative NOMA transmissions with energy harvesting (EH), secure communication is analyzed through AF EH relays in the presence of a passive eavesdropper (EVE) in \cite{Salem2020}. In \cite{Nguyen2024EHPNOMA}, the reliability and security of partial-NOMA systems based on EH are examined, showing that resource allocation and power-related parameters strongly affect both outage and secrecy performance. Wireless-powered cooperative NOMA systems under imperfect channel state information (CSI) and ipSIC are investigated in \cite{Ren2025Sensors}, where relay and jammer selection are used to improve secure transmission. The secrecy performance of a multi-antenna satellite-uncrewed aerial vehicle (UAV)-terrestrial SWIPT-NOMA system is analyzed, where the multi-antenna satellite employs maximum-ratio transmission (MRT), a single-antenna UAV relay assists the transmission, and the terrestrial users apply maximal-ratio combining (MRC) in the presence of a randomly located EVE and ipSIC~\cite{Zhang2024NOMASWIPTSecrecy}. These works show the importance of combining EH/SWIPT, cooperative transmission, and PLS in NOMA-based networks.

Artificial noise (AN) is an effective PLS technique in which an intentionally generated interference signal is used to degrade the eavesdropper channel \cite{Goel2008,Niu2026Survey}. In NOMA systems, AN can be generated by the transmitter, a cooperative user, a relay, or a dedicated jammer. However, AN generation also consumes power and may degrade the legitimate links if it is not properly designed or cancelled. Therefore, the power allocated to AN generation should be carefully selected, especially in SWIPT-assisted relaying where the harvested energy is shared between information forwarding and jamming.

Several AN-aided secure NOMA schemes have been reported in the literature. AN-aided secure beamforming for multi-input single-output (MISO)-NOMA systems is studied in \cite{Lv2018}, while the secrecy performance of massive multi-input multi-output (MIMO)-NOMA systems with AN is analyzed in \cite{Zeng2019}. In cooperative NOMA-based IoT networks, multiantenna-aided AN is used to improve the secrecy performance of legitimate IoT nodes \cite{Ruby2022}. AN-aided NOMA schemes have also been considered for UAV assisted and network-coded cooperative systems \cite{Cao2025,Yan2026}. Moreover, the effect of imperfect CSI and ipSIC on AN-assisted secure NOMA transmission is examined in \cite{Singh2020ImperfectCSI}. 
\textcolor{black}{In \cite{Cuong2024}, the SOP of a cooperative MIMO-NOMA relaying system employing MRT/MRC techniques is investigated in the presence of a multi-antenna EVE, where the effect of imperfect CSI at EVE is also considered.}
The authors in \cite{Le2019TASMRC} have studied an AN-aided MIMO-NOMA network with transmit antenna selection (TAS)/MRC over untrusted EH relays and derived analytical SOP expressions while examining the effects of parameters such as power allocation, power splitting (PS) ratio, and ipSIC.
Although these studies demonstrate the effectiveness of AN for enhancing PLS, the joint consideration of relay-generated AN, SWIPT-based AF relaying, diversity techniques such as orthogonal space-time block coding (OSTBC), receive antenna selection (RAS), MRC, and practical impairments such as channel estimation errors (CEE), feedback delay (FBD), and ipSIC has not been sufficiently addressed.

To clarify the place of this work with respect to the existing literature, Table~\ref{tab:literature_comparison} summarizes representative secrecy-oriented relaying NOMA studies. As shown in the table, existing works have addressed EH/SWIPT, AN-aided transmission, cooperative relaying, or multi-antenna secrecy from different perspectives. However, by remaining unadressed, the joint analysis of SWIPT-based AF relaying with relay-generated AN, OSTBC at the source, RAS at legitimate users, MRC at EVE, and practical impairments deserves to study.
\begin{table*}[t]
\centering
\caption{Comparison of representative secrecy-oriented relaying NOMA studies and the proposed system}
\label{tab:literature_comparison}
\renewcommand{\arraystretch}{1}
\scriptsize
\begin{tabular}{p{0.5cm}|p{3.2cm}|p{1cm}|p{1.5cm}|p{2.5cm}|p{2.5cm}|p{3.2cm}}
\hline
\textbf{Ref.} & \textbf{System model} & \textbf{EH/ SWIPT} & \textbf{AN/Jamming} & \textbf{Multi-antenna} & \textbf{Practical impairments} & \textbf{Performance metrics} \\
\hline
\cite{Lei2019} & Cooperative NOMA with relay selection & \large -- & \large -- & \large -- & \large-- & SOP \\
\hline
\cite{Salem2020} & EH cooperative NOMA with relay selection & \large\checkmark & \large-- & \large-- & \large-- & SOP \\
\hline
\cite{Nguyen2024EHPNOMA} & EH-based partial-NOMA system & \large\checkmark & \large\large-- & \large-- & \large-- & OP, SOP \\
\hline
\cite{Lv2018} & Secure MISO-NOMA downlink & \large-- & AN & Beamforming & ipSIC & SOP, secrecy diversity order \\
\hline
\cite{Zeng2019} & Massive MIMO-NOMA downlink & \large-- & AN & Massive MIMO & CEE & Ergodic secrecy rate, SEE \\
\hline
\cite{Ruby2022} & Cooperative NOMA-based IoT network & \large-- & AN & Multi-antenna S and IoT relay & \large-- & SOP, asymptotic SOP \\
\hline
\cite{Ren2025Sensors} & Wireless-powered cooperative NOMA with relay/jammer selection &\large \checkmark & Wireless-powered jammer &\large -- & Imperfect CSI, ipSIC & COP, SOP, EST, secrecy diversity order \\
\hline {\cite{Zhang2024NOMASWIPTSecrecy}} &
{NOMA-based multi-antenna satellite-UAV-terrestrial SWIPT system} &
{\large \checkmark} &
{\large --} &
{MRT at satellite, MRC at users and EVE} &
{ipSIC} &
{SOP, strictly positive secrecy capacity} \\
\hline
\cite{Le2019TASMRC} & MIMO/NOMA with multiple UEH relays & \large \checkmark & AN & TAS/MRC & ipSIC & SOP \\
\hline
\cite{Cuong2024} & Dual-hop MIMO-NOMA relay system & \large-- & \large-- & MRT/MRC, multi-antenna EVE & CEE at EVE & SOP \\
\hline
\textbf{This work} & \textbf{Dual-hop SWIPT-NOMA with AF relay-generated AN} &\large \checkmark &\large \checkmark & \textbf{OSTBC at source, RAS at users, MRC at EVE} & \textbf{CEE, FBD, ipSIC, imperfect AN cancellation} & \textbf{SOP, \textcolor{black}{secrecy diversity order}, secrecy throughput, SEE} \\
\hline
\end{tabular}
\end{table*}

\subsection{Motivation and Contributions}

The above discussion shows that PLS has been widely studied in NOMA, cooperative NOMA, and SWIPT-assisted NOMA networks. Existing studies have considered several secrecy-enhancing techniques, such as relay selection, jammer selection, AN-aided transmission, multi-antenna processing, and EH-based relaying. However, most of these works focus on only a subset of these aspects. In particular, the joint effect of relay-generated AN, SWIPT-based AF relaying, OSTBC transmission, RAS at legitimate users, MRC at EVE, and practical impairments has not been fully characterized in a unified secrecy analysis. Therefore, this paper considers a more general SWIPT-NOMA system, where the relay uses the harvested energy not only for forwarding the NOMA signal but also for generating AN to degrade the EVE link.

In the considered system, the source employs OSTBC transmission, legitimate users adopt RAS, and EVE applies MRC, which takes advantage of full benefit of receive diversity. The relay harvests energy from the received RF signal using the PS protocol and allocates the harvested energy between AF forwarding and AN generation. Moreover, CEE, FBD, ipSIC, and imperfect AN cancellation are incorporated into the analysis to reflect practical transmission conditions. The main contributions of this paper are summarized as follows:
\begin{itemize}
    \item A dual-hop AN-aided SWIPT-NOMA system is considered, where the source employs OSTBC transmission, legitimate users adopt RAS, and EVE applies MRC. Different from conventional cooperative NOMA models, the energy-constrained AF relay uses the harvested energy for both information forwarding and AN generation.

    \item The received signals and signal-to-interference+noise ratio (SINR) expressions are derived by jointly considering CEE, FBD, ipSIC, and imperfect AN cancellation.

    \item \textcolor{black}{Tractable analytical} SOP expressions are derived for the legitimate users over Nakagami-\(m\) fading channels. The analysis accounts for the statistical behavior of the OSTBC-based source-to-relay link, the RAS-based relay-to-user links, and the MRC-based relay-to-EVE link.

    \item \textcolor{black}{The high signal-to-noise ratio (SNR) secrecy behavior is characterized under both ideal \textcolor{black}{CSI} conditions and practical impairments. The ideal CSI analysis reveals the achievable secrecy diversity order, whereas the practical-impairment analysis demonstrates the existence of a nonzero secrecy outage floor.}

    \item Based on the derived SOP expressions, secrecy throughput and secrecy energy efficiency (SEE) are characterized. The numerical results verify the analytical expressions and demonstrate the effects of the OSTBC order, number of receive antennas, number of EVE antennas, PS ratio, AN power allocation factor and practical impairments on the system performance.
\end{itemize}

The rest of this paper is organized as follows. Section~II presents the system model and derives the received signals and SINR expressions. Section~III provides the SOP, \textcolor{black}{high-SNR secrecy analyses,} secrecy throughput, and SEE. The numerical results are discussed in Section~IV, and Section~V concludes the paper.

\section{System Model} \label{ChapVISystemModel} 

\begin{figure}[!t]
\centering
\includegraphics[width=1\linewidth]{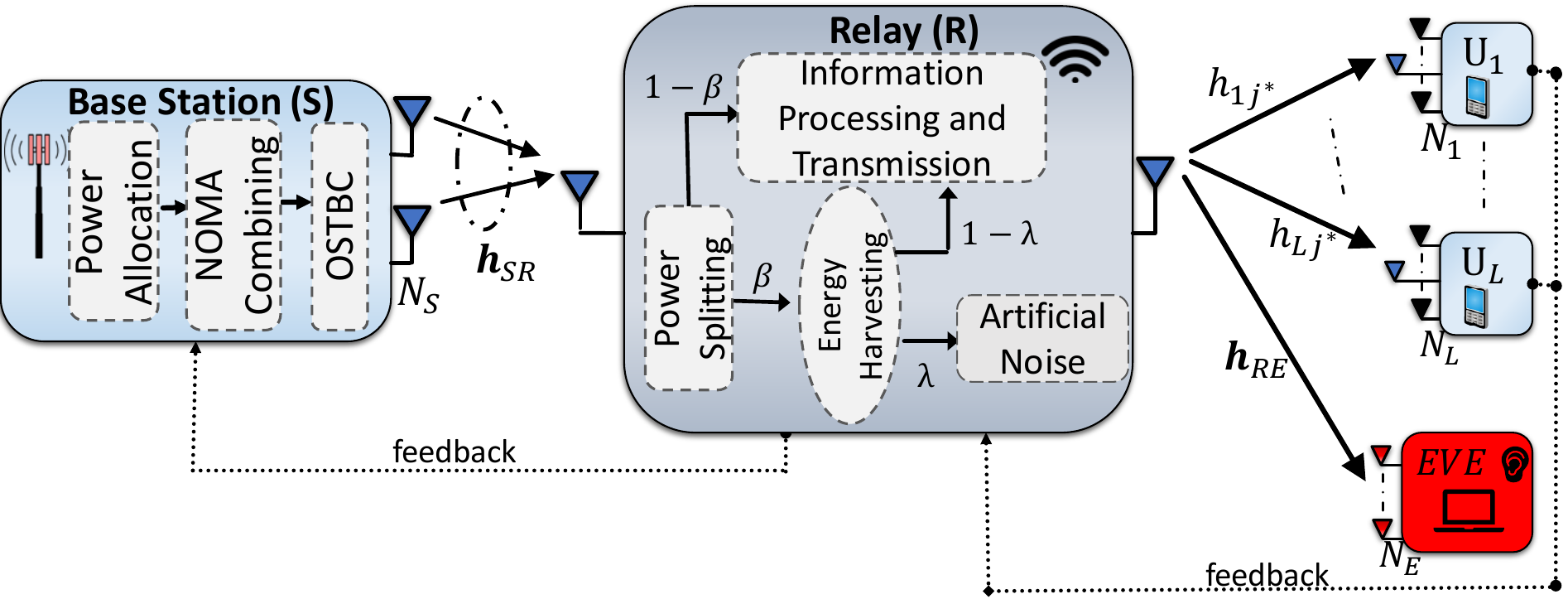}
\caption{Proposed dual-hop SWIPT-NOMA system with relay-generated AN, OSTBC at the S, RAS at users, and MRC at EVE.}
\label{fig1}

\end{figure}

We consider a cooperative downlink NOMA network consisting of a base station (S), a power-constrained relay (\(R\)), \(L\) legitimate users (\(U_l\)), and a passive EVE. It is assumed that direct links between the S and the terminal nodes are unavailable due to blockage or severe attenuation, and communication is established through a single-antenna relay operating in AF mode. 
\textcolor{black}{The S is equipped with \(N_S\) transmit antennas, each legitimate user is equipped with \(N_l\) receive antennas and employs RAS, whereas EVE is  equipped with \(N_E\) antennas and applies MRC.
At the S, the information symbols of the \(L\) users are superimposed in the power domain according to the NOMA principle. Specifically, the $k$-th NOMA symbol embedded into the OSTBC block is expressed as $s_k=\sum_{l=1}^{L}\sqrt{\frac{a_l P_S}{N_S R_c}}\,x_{l,k}$, where \(x_{l,k}\) denotes the unit-energy information symbol of \(U_l\), i.e., \(\mathbb{E}[|x_{l,k}|^2]=1\), \(P_S\) is the total transmit power of the S, and \(a_l\) is the power allocation coefficient satisfying $\sum_{l=1}^{L}a_l=1$. The power allocation coefficients are assigned according to the channel conditions of users, with higher power allocated to users with weaker channels. The resulting symbols are encoded using an OSTBC matrix $\mathbf{G}_{N_S}\in\mathbb{C}^{T\times N_S}$, where $K$ symbols are transmitted over $T$ time slots and $R_c=K/T$ denotes the OSTBC code rate, which becomes $1$ and $1/2$ for the $\mathbf{G}_2$ and $\mathbf{G}_3$ codes, respectively.}
The relay uses the harvested energy through the PS-SWIPT protocol to both forward the information signal and generate AN in order to degrade the reception quality at EVE. 

Before the data transmission phase, a training period is performed to obtain the CSI.
In the first hop, the relay estimates the channels from the \(N_S\) transmit antennas of the S to itself. 
In the second hop, each legitimate user estimates the channels between the relay and its \(N_l\) receive antennas.
\textcolor{black}{The available CSI is subsequently used for receive antenna selection, NOMA user ordering, and power allocation.} However, in practice, during the training period the channel coefficients are not perfectly estimated and thus may not exactly match the actual channel realizations during the data transmission phase. Besides, the feedback of the coefficients also introduces FBD. In order to characterize such real-world communication constraints, CEE, FBD, and in addition ipSIC are taken into account in the proposed system. 
\textcolor{black}{Following the channel model in \cite{Coskun2012}, the relationship between the actual channel coefficient during data transmission and the available outdated and imperfect channel estimate is modeled as $h_W = \rho_W \hat{h}_W^{\tau} + e_W, \qquad W \in \{SR,l\}$, where $\hat{h}_W^{\tau}$ denotes the outdated and imperfectly estimated channel coefficient and $\rho_W$ represents the correlation coefficient between the available CSI and the actual channel coefficient. The correlation coefficient is given by $\rho_W = J_0(2\pi f_D\tau)$, where $J_0(\cdot)$ is the zeroth-order Bessel function of the first kind, $f_D$ denotes the maximum Doppler frequency, and $\tau$ is the feedback delay. The aggregate channel error term $e_W$ accounts for both channel estimation and feedback-delay effects and is expressed as $e_W = e_{\mathrm{est},W}+e_{\mathrm{fbd},W}$, where $e_{\mathrm{est},W}\sim\mathcal{CN}\left(0,\sigma_{\mathrm{est},W}^{2}\right)$ and $e_{\mathrm{fbd},W}\sim\mathcal{CN}\left(0,\sigma_{\mathrm{fbd},W}^{2}\right)$. Accordingly, $e_W\sim\mathcal{CN}\left(0,\sigma_{e,W}^{2}\right)$, with $\sigma_{e,W}^{2}=\sigma_{\mathrm{est},W}^{2}+\sigma_{\mathrm{fbd},W}^{2}$, where $\sigma_{\mathrm{fbd},W}^{2} = \left(1-\rho_W^{2}\right)\hat{\Omega}_W$, and $\hat{\Omega}_W =\mathbb{E}\left[ \left|\hat{h}_W^{\tau}\right|^{2}\right]$. The estimation and feedback-delay error components are assumed to be statistically independent and uncorrelated with the available channel estimates.} Furthermore, a residual interference term is introduced during the SIC process in order to characterize the performance degradation caused by ipSIC.

In the first hop, the S transmits the superposed NOMA signal to the relay using the OSTBC scheme. Accordingly, the received signal at the relay is obtained as
\begin{equation}
	\label{Equation1}
         \textbf{y}_R = \mathbf{G}_{N_S} \textbf{h}_{\text{SR}} + \tilde{\textbf{w}}_R,
\end{equation}
where $\textbf{h}_{\text{SR}} \in \mathbb{C}^{N_S\times1}$ is the channel vector between S and relay, $\tilde{\textbf{w}}_R \in \mathbb{C}^{T\times1}$ is additive Gaussian noise (AGN). 
According to the PS-SWIPT protocol, the received signal at the relay is divided into two branches for information detection (ID) and EH. Specifically, a fraction \(1-\beta\) of the power of received signal is used for information processing, while the remaining fraction \(\beta\) is directed to the EH circuit, where \(0<\beta<1\). Thus, the signals at the ID and EH branches can be written as
$ \textbf{y}_{R}^{\mathrm{ID}} = \sqrt{1-\beta}\,  \textbf{y}_R = \sqrt{1-\beta}\ \mathbf{G}_{N_S} \textbf{h}_{\text{SR}} + {\textbf{w}}_R $ and $ \textbf{y}_{R}^{\mathrm{EH}} = \sqrt{\beta}\,  \textbf{y}_R$, respectively, where ${\textbf{w}}_R \in \mathbb{C}^{T\times1}$ whose entries follow $\mathcal{CN} (0, \sigma_{R}^2)$.
Based on the EH branch, the relay harvests energy from the received RF signal and uses this energy for transmission in the second hop. Let \(\eta\), \(0<\eta\leq 1\), denote the energy conversion efficiency. Then, the harvested power at the relay can be expressed as $P_R  = \eta \beta P_S \|\textbf{h}_{\text{SR}}\|^2 .$ In the second hop, the harvested power is divided into two parts. A fraction \(\lambda\) of \(P_R\) is allocated to AN generation, while the remaining fraction \(1-\lambda\) is used for the AF operation, where \(0\leq \lambda \leq 1\).
Therefore, the transmit signal at the relay consists of the amplified version of the ID-branch signal and the AN component generated by the remaining harvested power. Hence, the relay transmit signal is expressed as $\textbf{s}_R = \mathcal{G} \textbf{y}_{R}^{\mathrm{ID}} + \sqrt{\lambda P_R}\, \textbf{x}_{\mathrm{AN}}$, where \(\mathcal{G}\) denotes the amplifying factor and \(\mathbf x_{\rm AN}\in\mathbb C^{T\times1}\) is the AN signal vector satisfying \(\mathbb{E}\{|\textbf{x}_{\mathrm{AN}}|^2\}=1\). The amplifying factor is determined according to the average transmit power allocated to the AF operation. Accordingly, it can be written as $\mathcal{G} = \sqrt{\frac{(1-\lambda) P_R}{(1-\beta) P_S \|\mathbf{h}_{SR}\|^2 + N_S \sigma_R^2}}$. By substituting the harvested relay power \(P_R\) into the expression of $\mathcal{G}$ and after some algebraic manipulations, it can be rewritten as $\mathcal{G} \approx \sqrt{\frac{(1-\lambda) \eta \beta}{1-\beta}}$.

\textcolor{black}{In the second hop, the signal transmitted by the relay is received by both the legitimate users and EVE. Let $\hat{\tilde{h}}_{l,j}^{\tau}$ denote the available outdated and imperfect channel estimate corresponding to the link between the relay and the $j$-th receive antenna of $U_l$, where $j\in\{1,\ldots,N_l\}$ and $l\in\{1,\ldots,L\}$. Since the actual instantaneous channel coefficients are not perfectly available at the antenna-selection instant, RAS is performed based on the available CSI. Accordingly, the selected antenna index is given by $j_l^{\ast}=
\arg\max_{1\leq j\leq N_l}\left|\hat{\tilde{h}}_{l,j}^{\tau}\right|^{2}$. The corresponding selected channel estimate is denoted by $\hat{\tilde{h}}_{l}^{\tau} = \hat{\tilde{h}}_{l,j^{\ast}}^{\tau}$. The NOMA user ordering is also performed according to the available channel gains after RAS. Hence, the users are indexed such that $\left|\hat{h}_{1}^{\tau}\right|^{2} \leq \left|\hat{h}_{2}^{\tau}\right|^{2} \leq \cdots \leq \left|\hat{h}_{L}^{\tau}\right|^{2}$, where $\hat{h}_l^{\tau}$ denotes the ordered version of $\hat{\tilde{h}}_{l}^{\tau}$.
} Since the AN is generated by the relay, the legitimate users are assumed to know the AN signal. However, perfect AN cancellation is not possible in practice due to various non-ideal effects \cite{Guo2020}. Hence, only imperfect AN suppression is achieved at the legitimate users, and a residual AN term is retained in the received signal model. Let \(\zeta_l\), \(0 \leq \zeta_l \leq 1\), denote the residual AN coefficient for user \(U_l\), where \(\zeta_l=0\) corresponds to perfect AN cancellation and larger values of \(\zeta_l\) indicate more severe residual AN interference. After imperfect AN cancellation, the received signal at the selected antenna of \(U_l\) can be expressed as in \eqref{Equation3} at the top of the next page, \textcolor{black}{where $\mathbf{w}_l\in\mathbb{C}^{T\times1}$ is the AGN vector whose entries are distributed as $\mathcal{CN}(0,\sigma_l^2)$.} 

\begin{figure*}[!tp]
\small
\begin{equation}
\label{Equation3}
\begin{aligned}
\mathbf{y}_l
&=
\underbrace{
\sqrt{1-\beta}\,\mathcal{G}\,\rho_l \hat{h}_l^{\tau}
\rho_{SR}\hat{\mathbf{h}}_{SR}^{\tau}
\sqrt{a_l \frac{P_S}{N_S R_c}}\mathbf{x}_l
}_{\text{desired information component}}+
\underbrace{
\sqrt{1-\beta}\,\mathcal{G}\,\rho_l \hat{h}_l^{\tau}
\rho_{SR}\hat{\mathbf{h}}_{SR}^{\tau}
\sum_{\substack{i=1\\ i\neq l}}^{L}
\sqrt{a_i \frac{P_S}{N_S R_c}}\mathbf{x}_i
}_{\text{inter-user interference}}\\
&\quad+
\underbrace{
\sqrt{1-\beta}\,\mathcal{G}\,\mathbf{G}_{N_S}(\rho_l \hat{h}_l^{\tau}\mathbf{e}_{SR}
+\rho_{SR} e_l \hat{\mathbf{h}}_{SR}^{\tau}
+ e_l \mathbf{e}_{SR})}_{\text{CEE/FBD impairment terms}}
+
\underbrace{
\mathcal{G}\,\left(\rho_l \hat{h}_l^{\tau}
+ e_l\right) \mathbf{w}_R
}_{\text{amplified relay-noise term}}
+
\underbrace{
\sqrt{\zeta_l \lambda P_R}\,\rho_l \hat{h}_l^{\tau} \mathbf{x}_{\mathrm{AN}}
+\sqrt{\zeta_l \lambda P_R}\, e_l  \mathbf{x}_{\mathrm{AN}}
}_{\text{residual AN term}}
+
\underbrace{\mathbf{w}_l}_{\text{receiver noise}} .
\end{aligned}
\end{equation}
\hrulefill
\end{figure*}
Based on the above signal models, the received signal at each legitimate user consists of the desired information component, inter-user interference due to NOMA signaling, residual AN caused by imperfect AN cancellation, the amplified relay-noise term, the impairment terms due to CEE and FBD, and the receiver noise. Therefore, the instantaneous SINR for decoding the message of \(U_l\) can be generally expressed as

\begin{equation}
\label{Equation4}
\gamma_l = \frac{ \frac{\bar\gamma}{N_S R_c}a_lX Y_l} { \frac{\bar\gamma}{N_S R_c}\kappa_1 X Y_l + \bar\gamma \kappa_2 Y_l + \bar\gamma \kappa_3 X +  \bar\gamma \kappa_4 + \kappa_5 Y_l + \kappa_6},
\end{equation}
where $X = \| \hat{\mathbf{h}}_{SR}^{\tau}\|^2 $, $Y_l =\left|\hat{h}_l^{\tau} \right|^2 $, $\kappa_1 = \tilde{a}_l + \frac{R_c \zeta_l \lambda}{1-\lambda} $, $\kappa_2 = \left( 1+\frac{\zeta_l \lambda}{1-\lambda}\right) \frac{\sigma_{e,SR}^2}{\rho_{SR}^2}$, $\kappa_3 = \left( 1+\frac{\zeta_l \lambda}{N_S(1-\lambda)}\right) \frac{\sigma_{e,l}^2}{\rho_{l}^2}$, $\kappa_4 = \left( 1+\frac{\zeta_l \lambda}{1-\lambda}\right) \frac{\sigma_{e,SR}^2\sigma_{e,l}^2}{\rho_{SR}^2\rho_{l}^2}$, $\kappa_5 = \frac{1}{(1-\beta)\rho_{SR}^2}$, $\kappa_6 = \frac{1}{\rho_{SR}^2\rho_{l}^2} \left( \frac{\sigma_{e,l}^2}{1-\beta} + \frac{1}{\eta\beta(1-\lambda)}\right)$, $\bar{\gamma} = \frac{P_S}{\sigma_l^2}$, $\tilde{a}_l \triangleq \sum_{i=l+1}^{L} a_i + \sigma_{\mathrm{ipSIC}}^2 \sum_{i=1}^{l-1} a_i$ and $\sigma_{\mathrm{ipSIC}}^2$ denotes the normalized variance of the residual interference caused by ipSIC.

Finally, the signal transmitted by the relay is also received by EVE. Accordingly, the received signal vector at EVE can be written as
$\mathbf{y}_{E} = \mathbf{h}_{E} \textbf{s}_R + \mathbf{w}_{E}$,
where \(\mathbf{h}_{E} \in \mathbb{C}^{N_E \times 1}\) denotes the relay-to-EVE channel vector and \(\mathbf{w}_{E}  \in \mathbb{C}^{N_E\times1}\) is the AGN vector with independent entries distributed as \(\mathcal{CN}(0,\sigma_E^2)\). Since EVE is equipped with \(N_E\) antennas and employs MRC, the combined received signal can be expressed as
\begin{equation}
\label{Equation5}
\begin{aligned}
    \mathbf{y}_{E}&=\sqrt{1-\beta}\,\mathcal{G}\,\mathbf{G}_{N_S}(\rho_{SR}\hat{\mathbf{h}}_{SR}^{\tau} + \mathbf{e}_{SR})\|\mathbf{h}_{E}\|^2\\
    &+\mathcal{G}\,\|\mathbf{h}_{E}\|^2 \mathbf{w}_{R}+\sqrt{\lambda P_R}\,\textbf{x}_{\mathrm{AN}}\|\mathbf{h}_{E}\|^2+ \mathbf{h}_{E}^H \mathbf{w}_{E}.
\end{aligned}
\end{equation}
Unlike legitimate users, EVE does not have knowledge of the AN sequence generated at the relay. Therefore, the AN component acts as an additional interference term at EVE.
\textcolor{black}{In addition, EVE is assumed to follow the adopted NOMA decoding order and to perfectly cancel the messages decoded prior to $x_l$. Hence, when decoding $x_l$, the signals corresponding to $i<l$ are perfectly removed, whereas those corresponding to $i>l$ remain as inter-user interference. Accordingly, the aggregate interference coefficient at EVE is defined as $\tilde{a}_{E,l} \triangleq \sum_{i=l+1}^{L} a_i$.}
Consequently, the instantaneous SINR at EVE for decoding the message of \(U_l\) can be written as

\begin{equation}
\label{Equation6}
\gamma_{E,l} = \frac{ \frac{\bar\gamma_E}{N_S R_c}a_lX Z} { \frac{\bar\gamma_E}{N_S R_c}\kappa_{1e} X Z + \bar\gamma_E \kappa_{2e} Z + \kappa_{3e} Z + \kappa_{4e}},
\end{equation}
where $Z =\|\mathbf{h}_{E}\|^2$, $\kappa_{1e}  =\tilde{a}_{E,l} + \frac{R_c \lambda}{1-\lambda}$, $\kappa_{2e} = \frac{\sigma_{e,SR}^2}{(1-\lambda)\rho_{SR}^2}$, $\kappa_{3e} = \frac{\sigma_{R}^2}{(1-\beta)\rho_{SR}^2\sigma_{E}^2}$ and $\kappa_{4e} = \frac{1}{(1-\lambda)\eta\beta\rho_{SR}^2}$, $\bar{\gamma}_E = \frac{P_S}{\sigma_E^2}$ and $\tilde{a}_{E,l} \triangleq \sum_{i=l+1}^{L} a_i$. 

\section{Performance Analysis} \label{PerformanceSection} 

In this section, the secrecy performance of the considered PS-SWIPT-assisted cooperative NOMA system is analyzed in terms of SOP, \textcolor{black}{high-SNR secrecy behavior}, secrecy throughput, and SEE. 
To this end, by using the instantaneous SINR expressions derived in the previous section, an analytical SOP expression is first formulated.
\textcolor{black}{After that, the high-SNR secrecy behavior is investigated separately under ideal CSI conditions and practical impairments, yielding the asymptotic SOP and secrecy diversity order for the ideal CSI case and the secrecy outage floor for the practical case.}
Finally, the secrecy throughput and SEE metrics are presented based on the derived SOP expressions.

\subsection{SOP Analysis}
Based on the instantaneous SINR expressions of the legitimate user and EVE, the SOP expression of user \(U_l\) is derived in this subsection. For user \(U_l\), the instantaneous secrecy capacity is defined as the positive difference between the capacities of the main and wiretap channels. Accordingly, the secrecy capacity of \(U_l\) can be written as
$C_{s,l} = \left[\log_2(1+\gamma_l)-\log_2(1+\gamma_{E,l})\right]^+$,
where \([x]^+ = \max(x,0)\). Then, the SOP of user \(U_l\) is defined as
\begingroup
\begin{equation}
\label{Equation7}
\begin{aligned}
P_{\mathrm{S},l}
&= \Pr\left(C_{s,l} < R_{s,l}\right) = \Pr\left(\gamma_l < \psi_l(1+\gamma_{E,l}) - 1\right),
\end{aligned}
\end{equation}
\endgroup
where \(\psi_l = 2^{R_{s,l}}\) and $R_{s,l}$ is the secrecy target rate.
By using the definition in \eqref{Equation7}, the SOP of user \(U_l\) can be rewritten in terms of the cumulative distribution function (CDF) of \(\gamma_l\) and the probability density function (PDF) of \(\gamma_{E,l}\) as
\begingroup
\begin{equation}
\label{Equation8}
P_{\mathrm{S},l}
=
\int_{0}^{\infty}
F_{\gamma_l}\!\left(\psi_l(1+\gamma)-1\right)
f_{\gamma_{E,l}}(\gamma)\, d\gamma,
\end{equation}
\endgroup
where \(F_{\gamma_l}(\cdot)\) and \(f_{\gamma_{E,l}}(\cdot)\) denote the CDF of the instantaneous SINR at the legitimate user and the PDF of the instantaneous SINR at EVE, respectively.
By using \eqref{Equation4}, we first derive the CDF of the instantaneous SINR at the legitimate user as
\begin{equation}
\label{Equation9}
\begin{aligned}
F_{\gamma_l}(x)
&= \Pr\left(\gamma_l < x\right)\\ 
&= \Pr\left(
X <
\chi_l(x,y),
\;
Y_l>\frac{N_S R_c \kappa_3 x }{a_l-\kappa_1 x}
\right) \\
&= \int_{\frac{N_S R_c \kappa_3 x}{a_l-\kappa_1 x}}^{\infty}
F_X\!\left(\chi_l(x,y)\right)
f_{Y_l}(y)\,dy.
\end{aligned}
\end{equation}
which is valid for \(x<\frac{a_l}{\kappa_1}\). Otherwise, \(F_{\gamma_l}(x)=1\). Here $\chi_l(x,y)=\frac{N_S R_c x\left[By+(\bar{\gamma}\kappa_4+\kappa_6)\right]}{\bar{\gamma}\left[A_l(x)y-N_S R_c\kappa_3x\right]}$ with $A_l(x)=a_l-\kappa_1x$ and $B=\bar{\gamma}\kappa_2+\kappa_5$.
Then, by substituting the CDF of \(X\) and the PDF of \(Y_l\) into \eqref{Equation9}, the closed-form expression of \(F_{\gamma_l}(x)\) can be obtained after some algebraic manipulations as in \eqref{Equation10} at the top of next page,
\begin{figure*}[!tp]
\small
\begin{equation}
\label{Equation10}
\begin{aligned}
F_{\gamma_l}(x)
&=
1-
\frac{Q_l N_l \left(\dfrac{m_2}{\Omega_2}\right)^{m_2 N_l}}
{\Gamma(m_2)}
\sum_{k,p,q,r,s,n,t,v}C^{L-l}_{k}
C^{N_l-1}_{p}
C^{n}_{t}
C^{N_l(l+k-1)}_{r}C^{\phi_1-1}_{v}
\frac{(-1)^{k+p+r}}{n!}
\mu_q(p,m_2) \mu_s(r,m_2) 
\\
& \times\left(\frac{m_1N_S R_c}{\bar{\gamma}\Omega_1}\right)^n
B^t x^n
A_l(x)^{-(\phi_1-n)}
\left(N_S R_c\kappa_3 x\right)^{\phi_1-v-1} 
\exp\!\left[
-\left(
\frac{m_1 N_S R_c B}{\bar{\gamma}\Omega_1}
+
\frac{m_2(p+r+1)N_S R_c\kappa_3}{\Omega_2}
\right)
\frac{x}{A_l(x)}
\right]
\\
& \times D_l(x)^{n-t}
\left[
\frac{
m_1 N_S R_c x \Omega_2 D_l(x)
}{
\bar{\gamma}\Omega_1 m_2(p+r+1)
}
\right]^{\frac{\phi_2}{2}} 
K_{\phi_2}\!\left(
2\sqrt{
\frac{
m_1m_2N_S R_c(p+r+1)xD_l(x)
}{
\bar{\gamma}\Omega_1\Omega_2 A_l^2(x)
}
}
\right).
\end{aligned}
\end{equation}
\hrulefill
\end{figure*}
for \(x<\frac{a_l}{\kappa_1}\), whereas \(F_{\gamma_l}(x)=1\) otherwise. Here $\sum_{k,p,q,r,s,n,t,v}=\sum_{k=0}^{L-l}\sum_{p=0}^{N_l-1}\sum_{q=0}^{p(m_2-1)}\sum_{r=0}^{N_l(l+k-1)}$ $\sum_{s=0}^{r(m_2-1)}\sum_{n=0}^{m_1 N_S-1}\sum_{t=0}^{n}\sum_{v=0}^{\phi_1-1}$ and $\phi_1 = s+q+m_2$ and $\phi_2=t+v-n+1$. The detailed derivation is provided in Appendix~\ref{APP1}.

In order to evaluate \eqref{Equation8}, the PDF of the instantaneous SINR at EVE is needed. The closed-form expression of \(f_{\gamma_{E,l}}(x)\) is obtained as 

\begingroup
\small
\begin{equation}
\label{Equation11}
\begin{aligned}
f_{\gamma_{E,l}}(x)
&=
\frac{\left(\frac{m_E}{\Omega_E}\right)^{m_E N_E}}{\Gamma(m_E N_E)}
\sum_{t_1,t_2}
C^{t_1}_{t_2}
\frac{2\kappa_{4e}^{\,t_2}}{t_1!}\chi_E^{t_1}B_e^{t_1-t_2}\\
& \times
\frac{\left(\frac{\chi_E \Omega_E \kappa_{4e}}
{m_E}\right)^{\frac{\psi_E}{2}} a_l x^{\frac{\psi_E}{2}+t_1-1}}{\left(A_{e,l}(x)\right)^{\frac{\psi_E}{2}+t_1+1}} 
\exp\left(-\frac{\chi_E B_e x}{A_{e,l}(x)}\right) \\
& \times 
\Bigg\{
\Bigg[
\left(
\frac{\chi_EB_e x}{A_{e,l}(x)}
-\left(\frac{\psi_E}{2}+t_1\right)
\right)
K_{\psi_E}\!\bigl(\Delta_e(x)\bigr)
\\ &+\frac{\Delta_e(x)}{4}
\Bigl(
K_{\psi_E-1}\!\bigl(\Delta_e(x)\bigr)+
K_{\psi_E+1}\!\bigl(\Delta_e(x)\bigr)
\Bigr)
\Bigg]
\Bigg\},
\end{aligned}
\end{equation}
\endgroup
where $\psi_E = m_E N_E - t_2$, $\Delta_e(x)=2\sqrt{\frac{N_S R_c m_1 m_E \kappa_{4e}}{\bar{\gamma}_E \Omega_1 \Omega_E}\frac{x}{A_{e,l}(x)}}$, $\chi_E \triangleq \frac{N_S R_c m_1}{\bar{\gamma}_E \Omega_1}$ and 
$\sum_{t_1,t_2}=\sum_{t_1=0}^{m_1N_S-1}$ $\sum_{t_2=0}^{t_1}$.
For \(x \geq \frac{a_l}{\kappa_{1e}}\), it follows that \(f_{\gamma_{E,l}}(x)=0\). The detailed derivation is provided in Appendix~\ref{APP2}.

Finally, substituting \eqref{Equation10} and \eqref{Equation11} into \eqref{Equation8}, the integral-form SOP expression of user \(U_l\) is written as in \eqref{Equation12} at the top of next page. Here, \(\sum_{\mathcal{I}_l}\) denotes the multiple summation over the index set
$\mathcal{I}_l
=
\bigl\{
0\leq k\leq L-l,\;
0\leq p\leq N_l-1,\;
0\leq q\leq p(m_2-1),$ $
0\leq r\leq N_l(l+k-1),\;
0\leq s\leq r(m_2-1),$ $
0\leq n\leq m_1N_S-1,\;
0\leq t\leq n,\;
0\leq v\leq \phi_1-1,\;
0\leq t_1\leq m_1N_S-1,\;
0\leq t_2\leq t_1
\bigr\}$. Moreover, $\Phi_l(y)=\psi_l(1+y)-1,\qquad\Lambda_l(y)=a_l-\kappa_1\Phi_l(y)$,
and $\Xi_l(y)=B N_S R_c\kappa_3\Phi_l(y)+(\bar{\gamma}\kappa_4+\kappa_6)\Lambda_l(y)$. ${\varrho^*}=\min\left\{\frac{1}{\psi_l}\left(\frac{a_l}{\kappa_1}+1\right)-1,\frac{a_l}{\kappa_{1e}}\right\}$.
\begin{figure*}[!tp]
\begingroup
\small
\begin{equation}
\label{Equation12}
\begin{aligned}
P_{\mathrm{S},l}
&=
1-
\frac{
Q_l N_l a_l
\left(\dfrac{m_2}{\Omega_2}\right)^{m_2 N_l}
\left(\dfrac{m_E}{\Omega_E}\right)^{m_E N_E}
}
{\Gamma(m_2)\Gamma(m_E N_E)}
\sum_{\mathcal{I}_l}C^{L-l}_{k}
C^{N_l-1}_{p} C^{n}_{t}C^{t_1}_{t_2}
C^{N_l(l+k-1)}_{r}
C^{\phi_1-1}_{v}\frac{(-1)^{k+p+r}\kappa_{4e}^{\,t_2}}{n!t_1!} \\
&\times B^t\mu_q(p,m_2)\mu_s(r,m_2)
\left(\frac{m_1N_S R_c}{\Omega_1\bar{\gamma}}\right)^n
\chi_E^{t_1}B_e^{t_1-t_2}
\left(\frac{\chi_E\Omega_E\kappa_{4e}}
{m_E}\right)^{\frac{\psi_E}{2}}
\\
&\times \int_{0}^{{\varrho^*}}
\bigl(\Phi_l(y)\bigr)^n
\bigl(\Lambda_l(y)\bigr)^{-(\phi_1-n)}
\bigl(N_S R_c\kappa_3\Phi_l(y)\bigr)^{\phi_1-v-1}\exp\!\left(
-\frac{\chi_EB_e y}
{A_{e,l}(y)}
\right)
\\
&\times
\exp\!\left[
-\left(
\frac{m_1 N_S R_c B}{\bar{\gamma}\Omega_1}
+
\frac{m_2(p+r+1)N_S R_c\kappa_3}{\Omega_2}
\right)
\frac{\Phi_l(y)}{\Lambda_l(y)}
\right]
\\
&\times K_{\phi_2}\!\left(
2\sqrt{
\frac{
m_1m_2N_S R_c(p+r+1)\Phi_l(y)\Xi_l(y)
}{
\bar{\gamma}\Omega_1\Omega_2 \bigl(\Lambda_l(y)\bigr)^2
}
}
\right)
\frac{\bigl(\Xi_l(y)\bigr)^{n-t}
y^{\frac{\psi_E}{2}+t_1-1}
}
{
\bigl(A_{e,l}(y)\bigr)^{\frac{\psi_E}{2}+t_1+1}
}
\left[
\frac{
m_1 N_S R_c \Omega_2 \Phi_l(y)\Xi_l(y)
}{
\bar{\gamma}\Omega_1 m_2(p+r+1)
}
\right]^{\frac{\phi_2}{2}}
\\
&\times \Bigg\{
\Bigg[
\frac{\chi_EB_e y}
{A_{e,l}(y)}
-
\left(\frac{\psi_E}{2}+t_1\right)
\Bigg]
K_{\psi_E}\!\bigl(\Delta_e(y)\bigr)+
\frac{\Delta_e(y)}{4}
\left(
K_{\psi_E-1}\!\bigl(\Delta_e(y)\bigr)
+ K_{\psi_E+1}\!\bigl(\Delta_e(y)\bigr)
\right)
\Bigg\}
\,dy .
\end{aligned}
\end{equation}
\endgroup
\hrulefill
\end{figure*} 
Due to the mathematical complexity of the integral in \eqref{Equation12}, it is not straightforward to derive an exact closed-form expression for \(P_{\mathrm{S},l}\). Therefore, the Gaussian--Chebyshev quadrature technique is employed to obtain a tractable approximation. To this end, the variable transformation \( y=\frac{(u+1){\varrho^*}}{2} \) is applied to recast the integral into a form suitable for this quadrature rule.
Then, using the transformed integral together with the Gaussian--Chebyshev quadrature formula given in \cite[eq.(25.4.38)]{Abram1972}, the final approximate expression for \(P_{\mathrm{S},l}\) is obtained as
\begingroup
\small
\begin{equation}
\label{Equation13}
\begin{aligned}
P_{\mathrm{S},l}
&=
1-
\frac{
Q_l N_l a_l
\left(\dfrac{m_2}{\Omega_2}\right)^{m_2 N_l}
\left(\dfrac{m_E}{\Omega_E}\right)^{m_E N_E}
}
{\Gamma(m_2)\Gamma(m_E N_E)}
\sum_{\mathcal{I}_l}
C^{L-l}_{k} C^{N_l-1}_{p}\\
&\times C^{n}_{t} C^{N_l(l+k-1)}_{r}
C^{t_1}_{t_2}
C^{\phi_1-1}_{v}
\frac{(-1)^{k+p+r}\kappa_{4e}^{\,t_2}}{n!t_1!}
\left(\frac{m_1N_S R_c}{\Omega_1\bar{\gamma}}\right)^n
\\
&\times \mu_q(p,m_2)\mu_s(r,m_2) B^t
\chi_E^{t_1}
B_e^{t_1-t_2} \left(
\frac{\chi_E\kappa_{4e}}
{m_E}
\right)^{\frac{\psi_E}{2}}\\
&\times  \sum_{i=1}^{I} w_i f_{GC}(u_i).
\end{aligned}
\end{equation}
\endgroup
where \(u_i=\cos\!\left(\frac{(2i-1)\pi}{2I}\right)\) are the Chebyshev nodes, while \(w_i=\frac{\pi}{I}\) are the associated quadrature weights. The parameter \(I\) denotes the number of quadrature points used to approximate the integral. As \(I\) increases, the numerical approximation becomes more accurate at the cost of higher computational complexity.
Additionally, the function \(f_{GC}(u)\) is given as
\begingroup
\small
\begin{equation}
\label{Equation14}
\begin{aligned}
f_{GC}(u)
&=
\frac{{\varrho^*\sqrt{1-u^2}}}{2} \bigl(\Phi_l(y_u)\bigr)^n
\bigl(\Lambda_l(y_u)\bigr)^{-(\phi_1-n)}  \bigl(\Xi_l(y_u)\bigr)^{n-t}
\\
& \times \exp\!\left(
-\left[
\frac{m_1 N_S R_c B}{\bar{\gamma}\Omega_1}
+
\frac{m_2(p+r+1)N_S R_c\kappa_3}{\Omega_2}
\right]
\frac{\Phi_l(y_u)}{\Lambda_l(y_u)}
\right)
\\
&\times \bigl(N_S R_c\kappa_3\Phi_l(y_u)\bigr)^{\phi_1-v-1}
\left[
\frac{
m_1 N_S R_c \Omega_2 \Phi_l(y_u)\Xi_l(y_u)
}{
\bar{\gamma}\Omega_1 m_2(p+r+1)
}
\right]^{\frac{\phi_2}{2}}
\\
& \times K_{\phi_2}\!\left(
2\sqrt{
\frac{
m_1m_2N_S R_c(p+r+1)\Phi_l(y_u)\Xi_l(y_u)
}{
\bar{\gamma}\Omega_1\Omega_2\bigl(\Lambda_l(y_u)\bigr)^2
}
}
\right) 
\\
&\times \frac{y_u^{\frac{\psi_E}{2}+t_1-1}}{\bigl(A_{e,l}(y_u)\bigr)^{\frac{\psi_E}{2}+t_1+1}} \exp\!\left(
-\frac{\chi_E B_e y_u}
{A_{e,l}(y_u)}
\right)
\\
& \times
\Bigg\{
\Bigg[
\frac{\chi_E B_e y_u}
{A_{e,l}(y_u)}
-
\left(\frac{\psi_E}{2}+t_1\right)
\Bigg]
K_{\psi_E}\!\bigl(\Delta_e(y_u)\bigr)
\\&+\frac{\Delta_e(y_u)}{4}
\left[
K_{\psi_E-1}\!\bigl(\Delta_e(y_u)\bigr)
+
K_{\psi_E+1}\!\bigl(\Delta_e(y_u)\bigr)
\right]  \Bigg\}
\end{aligned}
\end{equation}
\endgroup

\subsection{\textcolor{black}{High-SNR Secrecy Performance Analysis}}
Although the obtained analytical SOP approximation accurately characterizes the secrecy performance of the considered system, their mathematical forms are still too involved to explicitly reveal the impact of the key system parameters. 
\textcolor{black}{Therefore, the high-SNR secrecy performance is investigated in this subsection to provide further insight into the dominant behavior of the considered system as \(\bar{\gamma}\rightarrow\infty\). In particular, an asymptotic SOP analysis is first carried out under ideal CSI conditions, from which the achievable secrecy diversity order is obtained. Subsequently, the high-SNR behavior under practical impairments is characterized in terms of the resulting secrecy outage floor.}

\subsubsection{Asymptotic SOP under Ideal \textcolor{black}{CSI} Conditions}

In this subsection, the asymptotic SOP expression is derived for the ideal \textcolor{black}{CSI} case in order to reveal the high-SNR secrecy behavior of the considered system in the absence of CEE and FBD. Under these assumptions, the parameters related to CEE and FBD vanish, i.e., \(\kappa_2=\kappa_3=\kappa_4=0\), while the effects of ipSIC and imperfect AN cancellation are still retained. 
\textcolor{black}{In this case, using the instantaneous SINR in (3), the CDF of \(\gamma_l\) can be expressed as  $F_{\gamma_l}(x) = \int_{0}^{\infty} F_X \left( \tau_l(x) \left( 1+\frac{\nu_l}{y} \right) \right) f_{Y_l}(y)\,dy$ for \(0<x<a_l/\kappa_1\), where $\tau_l(x) = \frac{N_SR_c\kappa_5x} {\bar{\gamma}(a_l-\kappa_1x)}, \qquad \nu_l=\frac{\kappa_6}{\kappa_5}.$ For \(x\geq a_l/\kappa_1\), \(F_{\gamma_l}(x)=1\). To obtain the high-SNR behavior of $F_{\gamma_l}(x)$, let $M=m_1N_S,  D_l=l m_2N_l$. As \(\bar{\gamma}\rightarrow\infty\), \(\tau_l(x)\rightarrow0\). Therefore, the asymptotic behavior of \(F_{\gamma_l}(x)\) is governed by the small-argument distributions of \(X\) and \(Y_l\). From the Gamma distribution of the equivalent S-to-R channel gain, $F_X(x) \sim A_Xx^M, x\rightarrow0$, where $A_X= \frac{1}{\Gamma(M+1)} \left( \frac{m_1}{\Omega_1} \right)^M.$ Similarly, by using the small-argument behavior of the RAS channel gain together with the \(l\)-th order statistic associated with NOMA user ordering, the CDF of \(Y_l\) behaves as $F_{Y_l}(y) \sim A_{Y_l}y^{D_l},y\rightarrow0,$ where $A_{Y_l} = \binom{L}{l} \left[ \frac{(m_2/\Omega_2)^{m_2}} {\Gamma(m_2+1)} \right]^{lN_l}.$ Accordingly, the dominant behavior of $F_{\gamma_l}(x)$ depends on the relative values of \(M\) and \(D_l\). By using the above small-argument representations and evaluating the corresponding negative moments, the asymptotic CDF of \(\gamma_l\) is obtained as
\begin{equation}
\small
\label{Equation15}
\begin{aligned}
F_{\gamma_l}^{\infty}(x)  &\approx  \begin{cases} \displaystyle    [\tau_l(x)]^M \Upsilon_{1,l} , & M<D_l, \\[5mm] \displaystyle [\tau_l(x)]^{D_l} \Upsilon_{2,l}, & M>D_l, \\[5mm] \displaystyle  [\tau_l(x)]^d \ln\!\left(\frac{1}{\tau_l(x)}\right) \Upsilon_{3,l}, & M=D_l=d . \end{cases}
\end{aligned}
\end{equation}
For  \(x\geq a_l/\kappa_1\), \(F_{\gamma_l}^{\infty}(x)=1\). The closed-form expressions of \(\Upsilon_{1,l}\), \(\Upsilon_{2,l}\), and \(\Upsilon_{3,l}\) are provided in Appendix~\ref{APP3}.}

\textcolor{black}{Then, by substituting \(F_{\gamma_l}^{\infty}(x)\) and \(f_{\gamma_{E,l}}(x)\) into \eqref{Equation8}, the asymptotic SOP expression under ideal CSI conditions can be expressed as}

\textcolor{black}{
\begin{equation}
\label{EquationXXasimp}
\begin{aligned}
P_{S,l}^{\infty}
&\approx
\Upsilon_l
\int_{0}^{\varrho^\star}
[\tau_l(\Phi_l(\gamma))]^{d_l^\star}
\left[
\ln\!\left(
\frac{1}{\tau_l(\Phi_l(\gamma))}
\right)
\right]^{\delta_l}
f_{\gamma_{E,l}}(\gamma)\,d\gamma ,
\end{aligned}
\end{equation}
where $d_l^\star=\min(M,D_l)$, $\delta_l=1$ for $M=D_l$ and
$0$ otherwise.}
\textcolor{black}{
The detailed derivation of \(P_{S,l}^{\infty}\) and its tractable Gaussian--Chebyshev representation are provided in Appendix~\ref{APP3}. Since the secrecy diversity order is determined by the dominant SNR dependence of the asymptotic SOP, it is sufficient here to examine the scaling of \(P_{S,l}^{\infty}\) with respect to \(\bar{\gamma}\).
Recall that $\tau_l(x) = \frac{N_SR_c\kappa_5x} {\bar{\gamma}(a_l-\kappa_1x)},$ and hence $ \tau_l(x)\propto\bar{\gamma}^{-1}.$ Moreover, \(\bar{\gamma}_E\) is assumed to remain fixed while \(\bar{\gamma}\rightarrow\infty\). Therefore, \(f_{\gamma_{E,l}}(\gamma)\) does not introduce any additional dependence on \(\bar{\gamma}\), and the EVE link affects only the SNR-independent coefficient of the asymptotic SOP. Accordingly, the dominant high-SNR behavior of the SOP is
\begin{equation}
\label{eq_asimp}
    P_{S,l}^{\infty} \propto \begin{cases} \bar{\gamma}^{-M}, & M<D_l, \\[1mm] \bar{\gamma}^{-D_l}, & M>D_l, \\[1mm] \bar{\gamma}^{-d}\ln\bar{\gamma}, & M=D_l=d. \end{cases}
\end{equation} 
Consequently, the secrecy diversity order of \(U_l\), defined as $ G_{d,l} = -\lim_{\bar{\gamma}\rightarrow\infty} \frac{\log P_{S,l}^{\infty}} {\log\bar{\gamma}},$ is obtained as
$ G_{d,l} = \min\left\{ M,\, D_l \right\} = \min\left\{ m_1N_S,\, lm_2N_l \right\}.$ }

\subsubsection{\textcolor{black}{Secrecy Outage Floor under Practical Impairments}}


\textcolor{black}{In the presence of practical impairments, the high-SNR secrecy performance is characterized by the limiting behavior of the instantaneous SINR at the legitimate user. Specifically, when \(\bar{\gamma}\rightarrow\infty\), the noise-related terms become negligible, whereas the interference terms associated with CEE and FBD remain. Accordingly, by dividing the numerator and denominator of \eqref{Equation4} by \(\bar{\gamma}\), the limiting SINR of \(U_l\) is obtained as}
\begingroup
\begin{equation}
\gamma_l^{\infty}
=
\frac{\frac{1}{N_S R_c}a_lXY_l}
{\frac{1}{N_S R_c}\kappa_1XY_l+\kappa_2Y_l+\kappa_3X + \kappa_4}.
\end{equation}
\endgroup
From the above SINR expression, it is readily observed that \(\gamma_l^{\infty}\) is independent of the transmit SNR \(\bar{\gamma}\). Consequently, the corresponding SOP converges to a nonzero constant as \(\bar{\gamma}\rightarrow\infty\), which reveals the existence of a secrecy outage floor under practical impairments, which implies zero diversity order.

\subsection{Secrecy Throughput Analysis}
In this subsection, the secrecy throughput of the considered system is analyzed based on the derived SOP expressions. Secrecy throughput is an important performance metric that quantifies the average successfully transmitted confidential rate by taking both the target secrecy rate and the SOP into account. For user \(U_l\), the secrecy throughput can be defined as $\mathcal{T}_l = \frac{1}{2}(1-P_{\mathrm{S},l})R_{s,l}$, where \(R_{s,l}\) denotes the target secrecy rate of user \(U_l\), and \(P_{\mathrm{S},l}\) is the corresponding SOP. Accordingly, the overall secrecy throughput of the considered \(L\)-user system can be expressed as
\begingroup
\begin{equation}
\label{EquationST2}
\mathcal{T}_{\mathrm{sum}}=\frac{1}{2}\sum_{l=1}^{L}(1-P_{\mathrm{S},l})R_{s,l}.
\end{equation}
\endgroup

\subsection{SEE Analysis}
In this subsection, the SEE of the considered system is investigated by taking into account both the secrecy throughput and the total consumed power. SEE is defined as the ratio of the achievable secrecy throughput to the total power consumption of the system. Accordingly, the SEE can be expressed as
\begingroup
\begin{equation}
\label{EquationEE1}
\eta_{\mathrm{EE}}=\frac{\mathcal{T}_{\mathrm{sum}}}{P_{\mathrm{total}}},
\end{equation}
\endgroup
where \(P_{\mathrm{total}} = P_S + P_C\) denotes the total consumed power of the considered system, \(P_S\) is the transmit power of the source, and \(P_C\) represents the total circuit power consumption of the system and is calculated according to the power consumption model in \cite{Cui2004}. By substituting \eqref{EquationST2} into \eqref{EquationEE1}, the SEE can be directly evaluated once the SOP values are obtained. This metric provides useful insight into the trade-off between secure transmission performance and energy consumption. In particular, although increasing the transmit power may improve the secrecy throughput in some operating regimes, the corresponding increase in total power consumption may limit the achievable SEE.

\section{Numerical Results} \label{ChapVINumericalResults}

In this section, numerical and simulation results are presented to validate the accuracy of the derived analytical expressions and to provide further insight into the secrecy performance of the considered system. All simulation curves are obtained in MATLAB and are averaged over \(10^7\) independent channel realizations. Unless otherwise stated, the energy conversion efficiency and Nakagami-\(m\) fading parameter are set to \(\eta=0.7\), and \(m=1\), respectively. The power allocation coefficients and target data rates are chosen as \((a_1, a_2, a_3)=(0.5, 0.3, 0.2)\) and \((R_{s,1}, R_{s,2}, R_{s,3})=(0.2, 0.2, 0.2)\), respectively. The channel gains are stated as $\Omega_1=\Omega_2=\Omega_E = 1$. The number of receive antennas at users and EVE are $N_l=N_E=2$. The PS ratio and the AN power allocation factor are $\beta=0.5$ and $\lambda=0.2$, respectively. For the ideal \textcolor{black}{CSI} case, perfect \textcolor{black}{channel estimation} and zero feedback delay are assumed, i.e., \(\sigma_{e,\mathcal{W}}^2=0\) and \(f_D\tau=0\). For the practical \textcolor{black}{CSI} case, the CEE variance and normalized delay parameter are set to \(\sigma_{e,\mathcal{W}}^2=0.03\) and \(f_D\tau=0.03\), respectively. In numerical evaluations, the Gaussian--Chebyshev quadrature is implemented with \(I=100\) nodes to obtain stable and sufficiently accurate analytical results.

\begin{figure}
    \centering
	\includegraphics[width=3.1in]{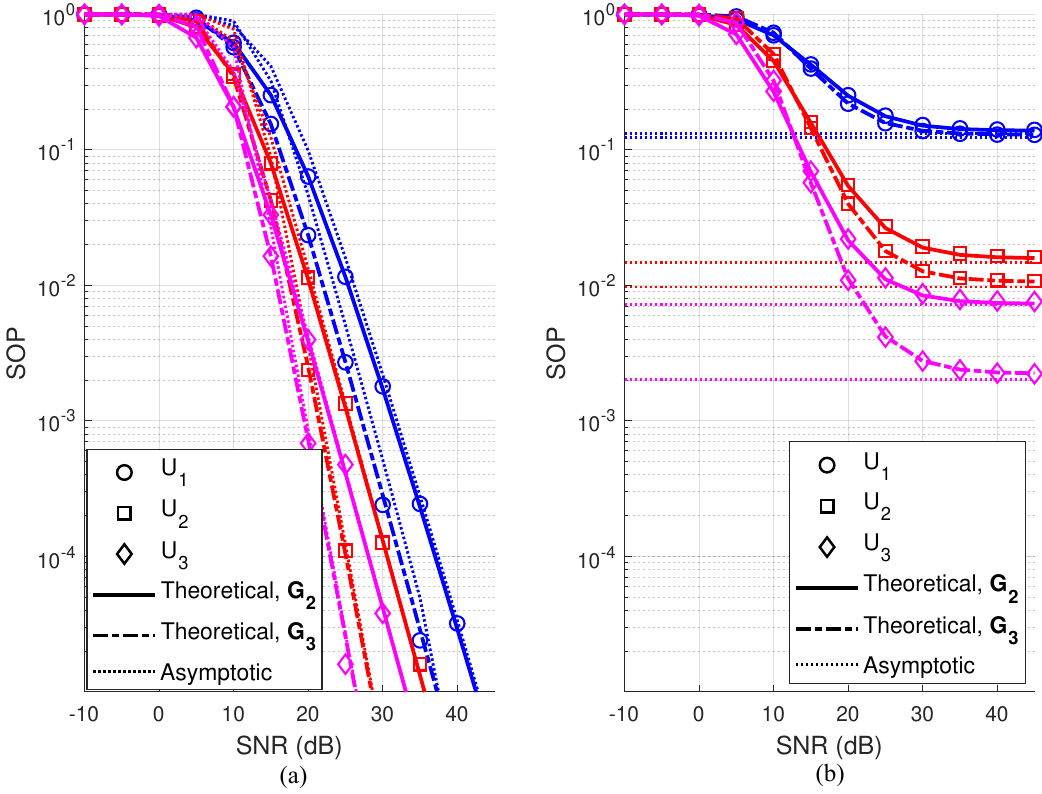}
	\caption{SOP performance of users versus transmit SNR for the \(\mathbf{G}_2\) and \(\mathbf{G}_3\) OSTBC structures under (a) ideal \textcolor{black}{CSI} conditions and (b) practical \textcolor{black}{CSI} conditions.}
	\label{fig2}
\end{figure}

Fig.~\ref{fig2} compares the SOP performance of the considered system for the \(\mathbf{G}_2\) and \(\mathbf{G}_3\) OSTBC structures under ideal and practical \textcolor{black}{CSI} conditions, where $\sigma_{ipSIC}^2 = 0$, $\lambda=0.2$ and $\zeta_l=0.01$. In Fig.~\ref{fig2}(a), CEE and FBD are neglected. For all users, the SOP decreases as the transmit SNR increases, since the legitimate links become more reliable at higher SNR values. It is also observed that \(\mathbf{G}_3\) outperforms \(\mathbf{G}_2\) due to the additional transmit diversity. For a target SOP of \(10^{-4}\), the SNR gains achieved by \(\mathbf{G}_3\) over \(\mathbf{G}_2\) are approximately \(4.75\), \(5.52\), and \(5.13\) dB for \(U_1\), \(U_2\), and \(U_3\), respectively. The asymptotic curves agree with the exact results in the high-SNR region. 
\textcolor{black}{In particular, for \(m=1\) and \(N_l=2\), the diversity orders are \(2\) for all users with \(\mathbf{G}_2\), whereas they are \(2\), \(3\), and \(3\) for \(U_1\), \(U_2\), and \(U_3\), respectively, with \(\mathbf{G}_3\)}, which is consistent with the slopes of the SOP curves and confirms the diversity improvement obtained by increasing the OSTBC order.
Fig.~\ref{fig2}(b) shows the results for the practical case. In contrast to the ideal \textcolor{black}{CSI} case, the SOP curves approach non-zero error floors at high SNR values. Therefore, the $G_{d,l}$ cannot be interpreted in the usual way under practical \textcolor{black}{CSI} conditions. Nevertheless, \(\mathbf{G}_3\) still provides a noticeable diversity-related performance advantage in the high SNR region.

\begin{figure}
	\centering
	\includegraphics[width=3.05in]{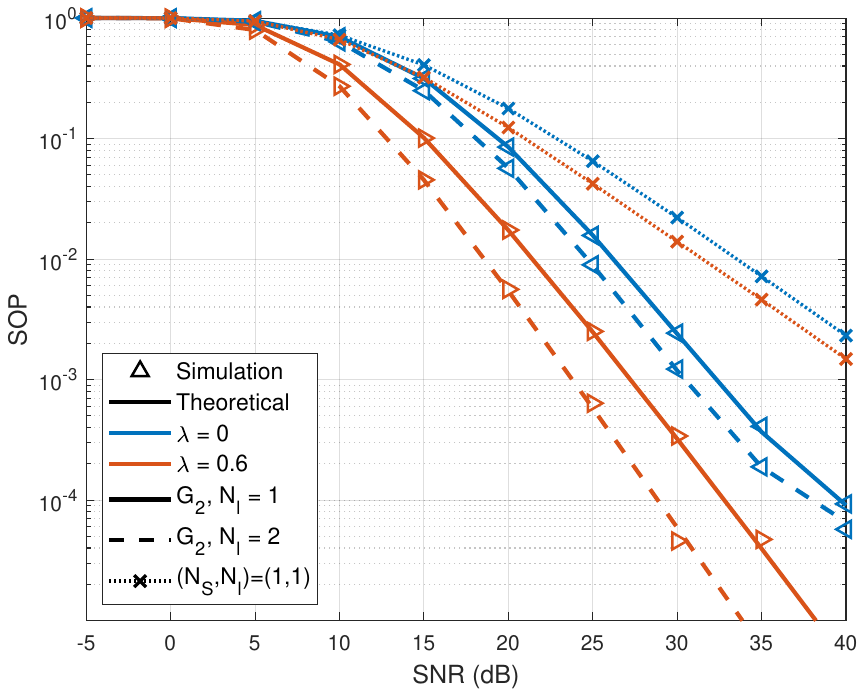}
	\caption{SOP of \(U_2\) versus transmit SNR for different \((N_S,N_l)\) configurations and AN power allocation factors.}
    \label{fig10}
\end{figure}

Fig.~\ref{fig10} illustrates the individual and joint benefits of OSTBC, RAS, and AN under ideal \textcolor{black}{CSI} conditions by comparing the benchmark \((N_S,N_l)=(1,1)\) case with the \(\mathbf{G}_2\) structure for \(N_l=1\) and \(N_l=2\), where \(N_E=2\), \(\sigma_{\mathrm{ipSIC}}^2=0\), \(\beta=0.8\), and \(\lambda \in \{0,0.6\}\). It is observed that the benchmark case gives the highest SOP, while employing \(\mathbf{G}_2\) improves the secrecy performance through transmit diversity. A further SOP reduction is observed when \(N_l\) increases from 1 to 2, indicating that RAS enhances the secrecy performance by improving the quality of the legitimate link. In addition, the curves with \(\lambda=0.6\) remain below those with \(\lambda=0\), showing the positive impact of relay-generated AN. For example, at a SOP of \(SOP=10^{-3}\), the SNR gain obtained by increasing \(\lambda\) from 0 to 0.6 for \(U_2\) is about \(5.10\) dB when \(N_l=1\), whereas it increases to about \(6.68\) dB when \(N_l=2\). Therefore, the joint use of OSTBC, RAS, and AN provides the most favorable secrecy performance.

\begin{figure}
	\centering
	\includegraphics[width=3.1in]{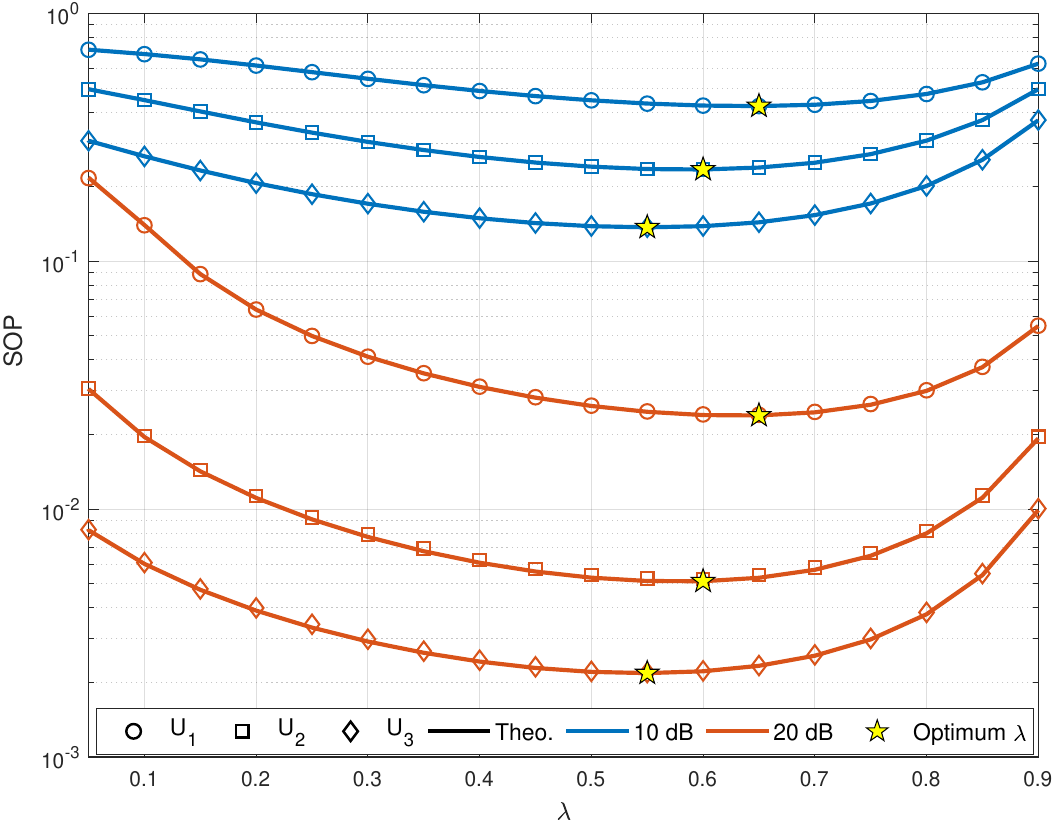}
	\caption{SOP versus AN power allocation factor \(\lambda\) for users under ideal \textcolor{black}{CSI} conditions with perfect SIC.}
    \label{fig3}
\end{figure}

Fig.~\ref{fig3} shows the SOP vs. the AN power allocation factor \(\lambda\) for users when \((N_S,N_l,N_E)=(2,2,2)\). The results are obtained under ideal \textcolor{black}{CSI} conditions and perfect SIC. For all users, the SOP first decreases as \(\lambda\) increases, reaches a minimum value, and then starts to increase. This trend is due to the power sharing at the relay. When \(\lambda\) is small, the AN power is not sufficient to effectively degrade the EVE link. On the other hand, when \(\lambda\) becomes large, less harvested power is allocated to AF transmission, which weakens the legitimate links and increases the SOP. Therefore, an optimum \(\lambda\) exists for each user. For the considered parameter set, the optimum value of \(\lambda\) is around \(0.6\) for all users.  It is also seen that increasing the transmit SNR from \(10\) dB to \(20\) dB improves the SOP performance, while the same trade-off with respect to \(\lambda\) is preserved. Similar trends are also observed for other antenna configurations. However they are not included in the figure to avoid overcrowding and to keep the presentation clear.

\begin{figure}
	\centering
	\includegraphics[width=3.1in]{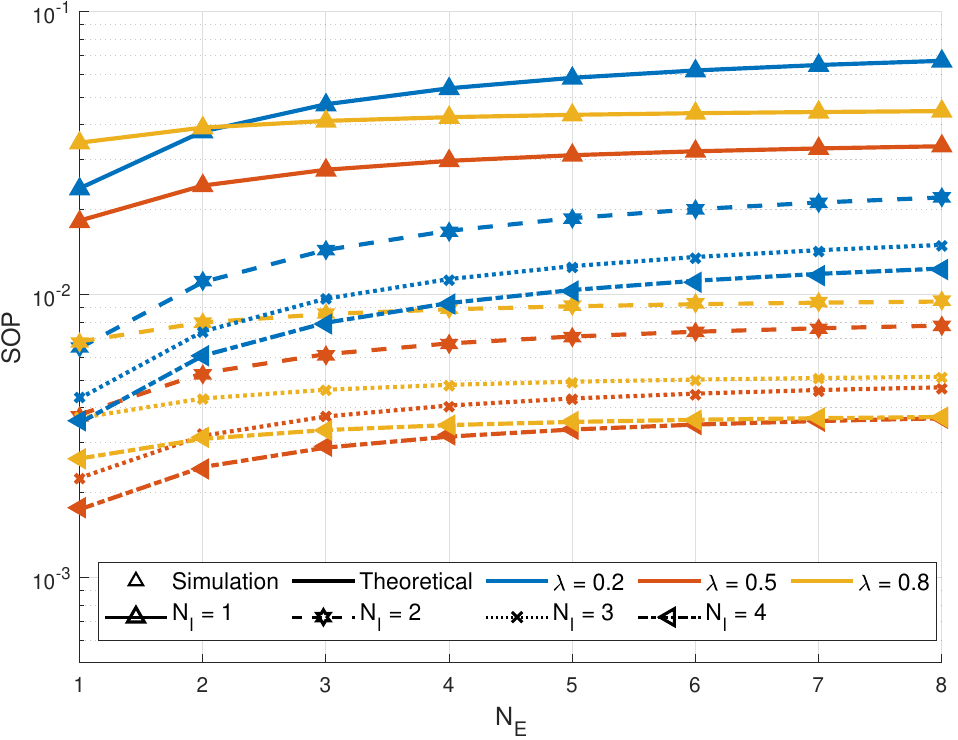}
    \caption{SOP of \(U_2\) versus the number of EVE antennas \(N_E\) for different values of \(N_l\) and \(\lambda\).}
    \label{fig4}
\end{figure}

Fig.~\ref{fig4} presents the SOP of \(U_2\) as a function of the number of EVE antennas \(N_E\) for different \(N_l\) and \(\lambda\) values, with \(N_S=2\), \(\beta=0.5\), and SNR \(=20\) dB. Only \(U_2\) is shown for clarity, since the other users exhibit similar behavior. The SOP increases as \(N_E\) grows, which is due to the MRC gain at EVE. On the other hand, increasing \(N_l\) improves the secrecy performance through RAS. This can be seen from the fact that, for each \(\lambda\), the curves with larger \(N_l\) stay below those with smaller \(N_l\). Hence, the receive diversity at the legitimate user partly mitigates the adverse effect of increasing \(N_E\). 
The figure also shows that the AN power allocation factor \(\lambda\) has a direct impact on the SOP. Particularly, for properly selected \(\lambda\) values, the SOP degradation caused by increasing \(N_E\) becomes less pronounced, indicating that relay-generated AN can effectively suppress the EVE diversity gain. Nevertheless, allocating excessive power to AN generation reduces the power available for AF transmission. Thus, the best secrecy performance is obtained by jointly considering the receive diversity gain and the AN power allocation.

\begin{figure}
	\centering
	\includegraphics[width=3.35in]{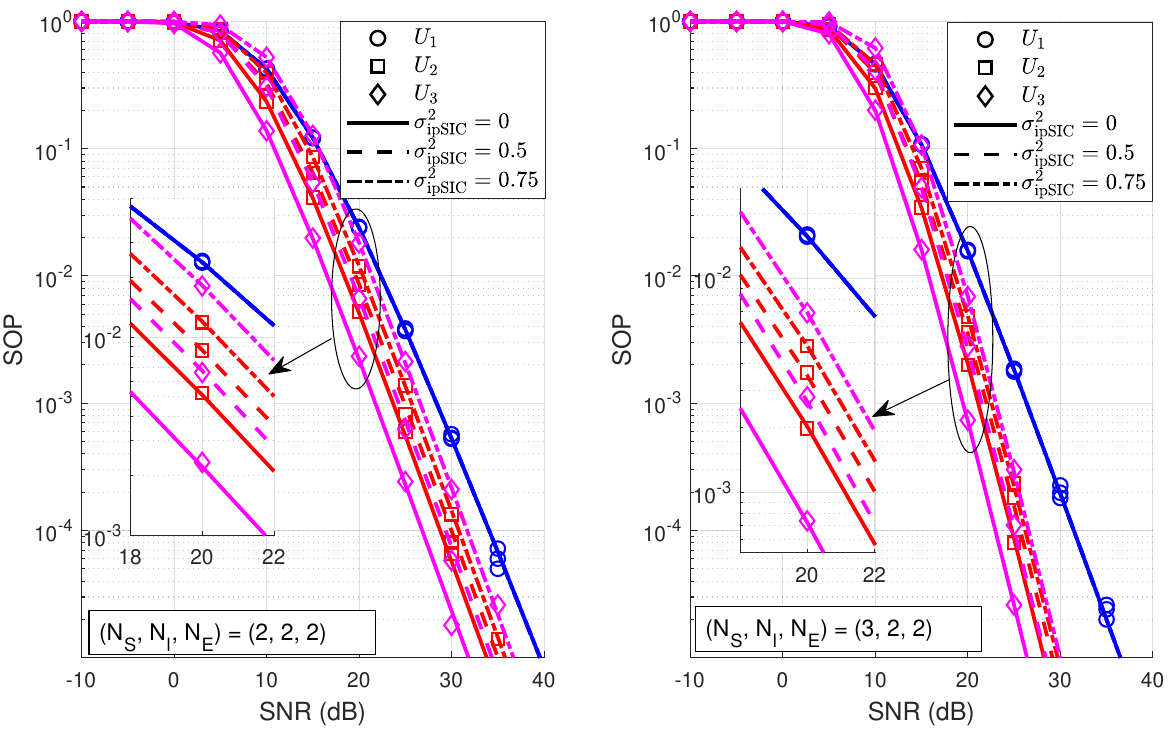}
    \caption{SOP performance of the users versus transmit SNR for different ipSIC levels under ideal \textcolor{black}{CSI} conditions: (a) \((N_S,N_l,N_E)=(2,2,2)\) and (b) \((N_S,N_l,N_E)=(3,2,2)\).}
    \label{fig5}
\end{figure}

Fig.~\ref{fig5} shows the SOP versus transmit SNR for different ipSIC levels under ideal \textcolor{black}{CSI} conditions, where \(\lambda=0.2\), \(\zeta_l=0.01\) and \(\beta=0.5\). The results are given for \((N_S,N_l,N_E)=(2,2,2)\) in Fig.~\ref{fig5}(a) and \((N_S,N_l,N_E)=(3,2,2)\) in Fig.~\ref{fig5}(b). Since \(U_1\) does not perform SIC, its performance is not affected by \(\sigma_{\mathrm{ipSIC}}^2\). On the other hand, \(U_2\) and \(U_3\) are degraded by the residual interference caused by ipSIC, with \(U_3\) being more sensitive due to the higher number of SIC operations. The best SOP performance is obtained for a perfect SIC, i.e., \(\sigma_{\mathrm{ipSIC}}^2=0\). As \(\sigma_{\mathrm{ipSIC}}^2\) increases, the SOP performance gradually degrades, especially in the moderate and high-SNR regions, due to residual interference introduced by imperfect SIC. The enlarged views in the insets more clearly highlight the performance differences among the considered ipSIC levels in the high-SNR region. It is also observed that increasing \(N_S\) shifts the curves to lower SOP values due to the transmit diversity gain. Similar trends are obtained for other antenna configurations, but they are not included to avoid visual complexity.

\begin{figure}
	\centering
	\includegraphics[width=3.2in]{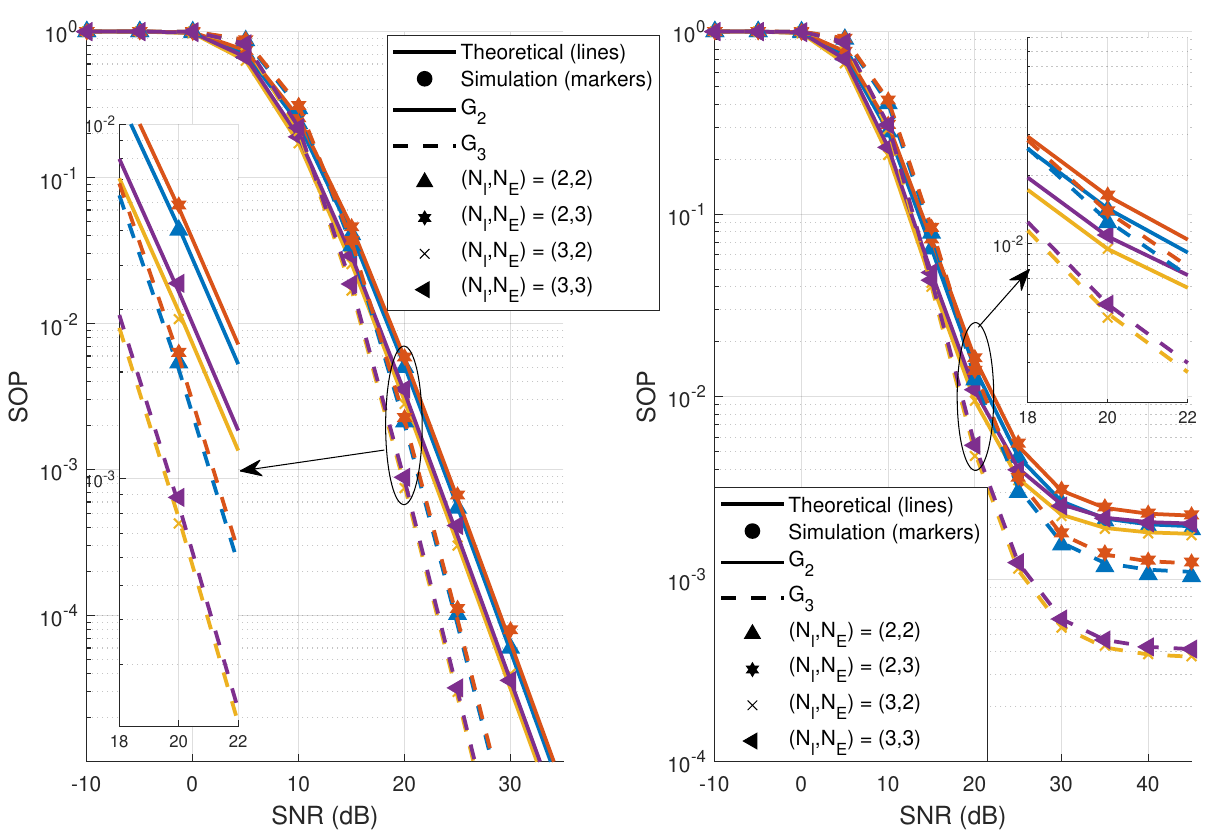}
    \caption{SOP performance of \(U_2\) versus transmit SNR for the \(\mathbf{G}_2\) and \(\mathbf{G}_3\) OSTBC structures under different \((N_l,N_E)\) configurations: (a) ideal \textcolor{black}{CSI} conditions and (b) practical \textcolor{black}{CSI} conditions.}
    \label{fig6}
\end{figure}

Fig.~\ref{fig6} presents the SOP performance of \(U_2\) versus the transmit SNR for the \(\mathbf{G}_2\) and \(\mathbf{G}_3\) OSTBC structures under different \((N_l,N_E)\) configurations. The results are obtained for \(\lambda=0.6\), \(\beta=0.5\), \(\zeta_l=0.01\), and \(\sigma_{\mathrm{ipSIC}}^2=0\). Fig.~\ref{fig6}(a) and Fig.~\ref{fig6}(b) correspond to the ideal and practical \textcolor{black}{CSI} cases, respectively. In both subfigures, \(\mathbf{G}_3\) provides lower SOP values than \(\mathbf{G}_2\) for all antenna configurations. It is observed that increasing \(N_l\) improves the SOP performance, whereas increasing \(N_E\) degrades it due to the stronger combining capability of EVE. Consequently, among the considered cases, \((N_l,N_E)=(3,2)\) provides the best performance, while \((N_l,N_E)=(2,3)\) gives the worst. However, for the optimum AN power allocation factor \(\lambda=0.6\), the difference between the \((2,2)\) and \((2,3)\) cases remains very small, indicating that the effect of one additional EVE antenna can be largely compensated by the adopted AN allocation. Moreover, the enlarged views in Fig.~\ref{fig6} show that the \(\mathbf{G}_3\) scheme suppresses the SOP degradation caused by increasing \(N_E\) more effectively than the \(\mathbf{G}_2\) scheme. Comparing Fig.~\ref{fig6}(a) and Fig.~\ref{fig6}(b), the same overall trends are preserved under practical conditions, although the SOP performance degrades due to the considered impairments. To keep the figure readable, only the results for \(U_2\) are presented, as the other users exhibit similar behavior.

\begin{figure}
	\centering
	\includegraphics[width=3in]{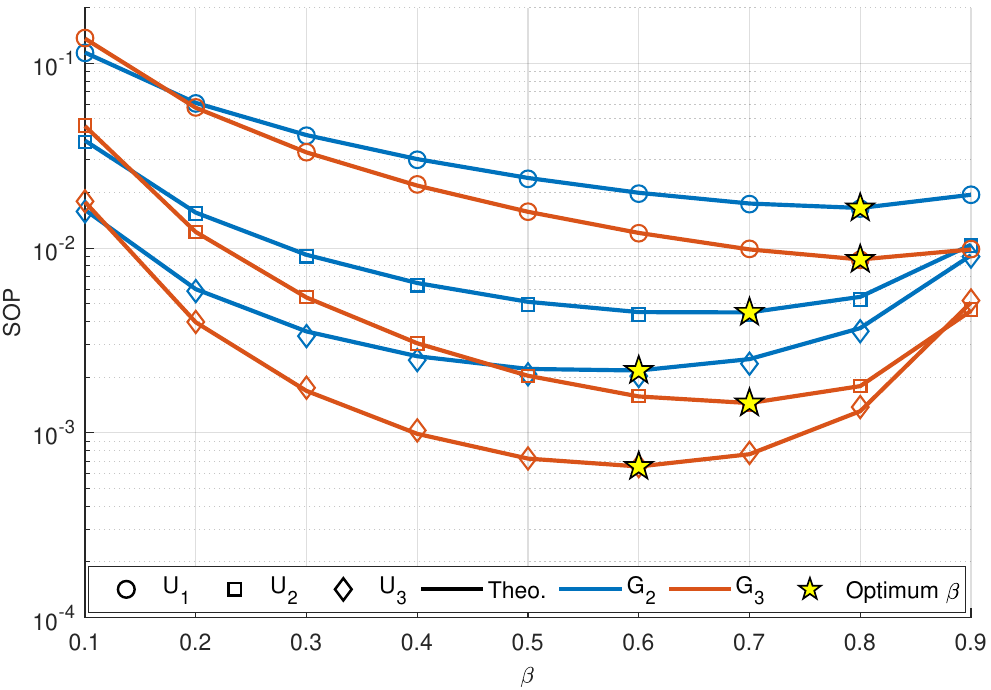}
    \caption{SOP versus the PS ratio \(\beta\) for users with the \(\mathbf{G}_2\) and \(\mathbf{G}_3\) OSTBC structures under ideal \textcolor{black}{CSI} conditions.}
    \label{fig7}
\end{figure}

Fig.~\ref{fig7} shows the SOP versus the PS ratio \(\beta\) for users under ideal \textcolor{black}{CSI} conditions, where \(\lambda=0.6\), \(\mathrm{SNR}=20\) dB, \(\zeta_l=0.01\), and \(\sigma_{\mathrm{ipSIC}}^2=0\). The results are presented for \((N_S,N_l,N_E)=(2,2,2)\) and \((3,2,2)\). It is observed that the SOP first decreases with \(\beta\), reaches a minimum point, and then increases. This behavior is due to the trade-off introduced by the PS at the relay. When \(\beta\) is small, the harvested energy is limited, and the relay cannot transmit the signal efficiently. On the other hand, when \(\beta\) becomes too large, less power remains for information processing, which also degrades the secrecy performance. Therefore, an optimum \(\beta\) exists for each user, as indicated in the figure. It is also seen that \(\mathbf{G}_3\) achieves lower SOP values than \(\mathbf{G}_2\) for both antenna configurations, showing the benefit of the higher transmit diversity order. In other words, the \((3,2,2)\) configuration provides better performance than \((2,2,2)\), which is consistent with the improvement obtained by increasing the number of transmit antennas. It is also observed that the optimum \(\beta\) decreases as the user index increases. This is because higher-order users are allocated lower power coefficients and additionally perform SIC, making their SINR more sensitive to the reduction in information-processing power caused by large \(\beta\) values. Consequently, these users achieve their optimum secrecy performance in smaller PS ratios. Moreover, nearly identical optimum \(\beta\) values are obtained for both \(\mathbf{G}_2\) and \(\mathbf{G}_3\). Similar trends are observed for other antenna configurations; however, they are not included to keep the figure clear.

\begin{figure}
	\centering
	\includegraphics[width=3.3in]{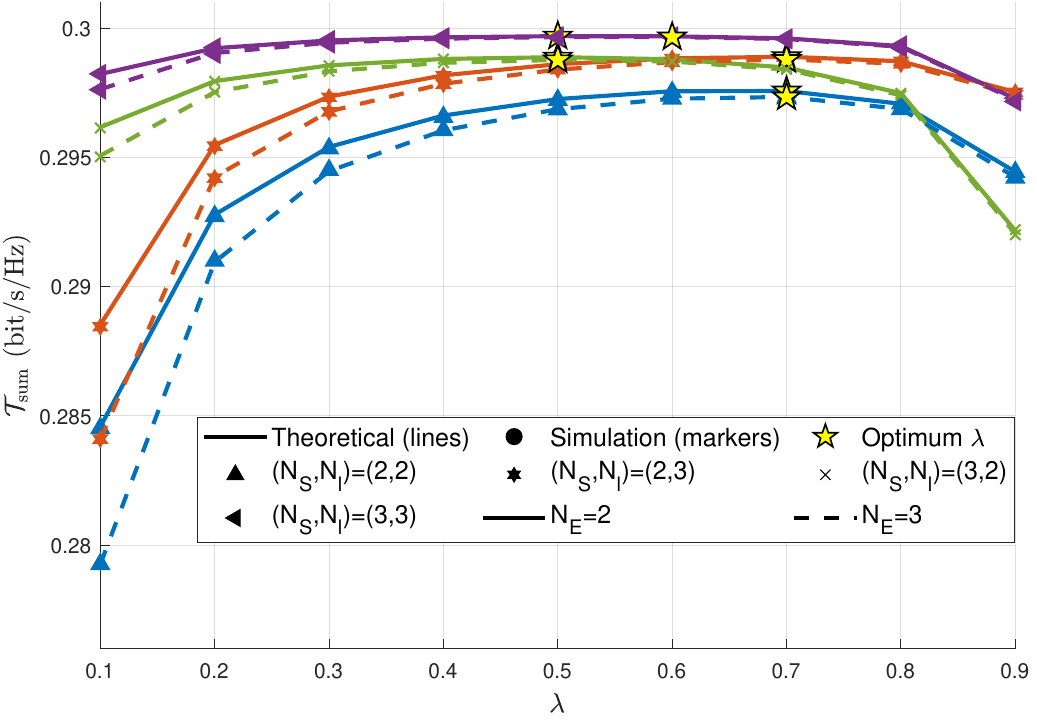}
    \caption{Sum secrecy throughput versus AN power allocation factor \(\lambda\) for different \((N_S,N_l)\) and \(N_E\) configurations.}
    \label{fig8}
\end{figure}

\begin{figure}
	\centering
	\includegraphics[width=3.3in]{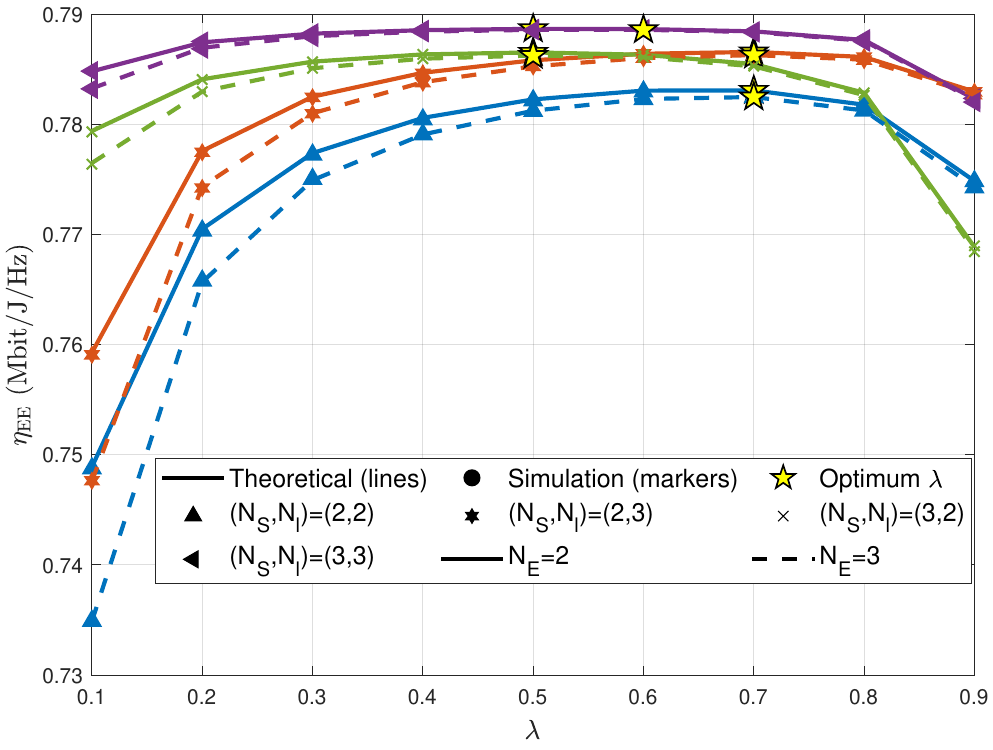}
    \caption{SEE versus AN power allocation factor \(\lambda\) for different \((N_S,N_l)\) and \(N_E\) configurations.}
    \label{fig9}
\end{figure}

Figs.~\ref{fig8} and \ref{fig9} show the sum secrecy throughput and SEE versus the AN power allocation factor \(\lambda\), respectively. The results are obtained for different configurations \((N_S, N_l, N_E)\), where \(\beta=0.7\), \(\zeta_l=0.01\), and SNR \(=20\) dB. For SEE, the circuit power consumption is calculated as \(P_C=P_{c,S}+P_{c,R}+L P_{c,U}+P_{c,\mathrm{AN}}\), where \(P_{c,S}=100\) mW, \(P_{c,R}=120\) mW, \(P_{c,U}=50\) mW, and \(P_{c,\mathrm{AN}}=10\) mW are used. In both figures, the performance first improves as \(\lambda\) increases and then degrades after a certain point. This behavior follows from the power allocation at the relay. For small \(\lambda\), the AN power is not sufficient to effectively limit the EVE link. As \(\lambda\) increases, the secrecy performance improves due to the stronger AN component. However, when \(\lambda\) becomes too large, the power allocated to AF transmission decreases, which weakens the legitimate links and reduces both the sum secrecy throughput and SEE. Hence, there is an optimum \(\lambda\) for each antenna configuration. It is also observed that increasing \(N_S\) or \(N_l\) improves the performance due to the additional transmit and receive diversity gains. Although a larger \(N_E\) generally strengthens the EVE link and degrades the secrecy performance, this effect becomes very limited around the optimum \(\lambda\) region. In particular, the curves corresponding to \(N_E=2\) and \(N_E=3\) are almost overlapping near their maximum points. This indicates that a proper allocation of AN power can effectively compensate for the performance loss caused by the additional EVE antenna. The optimum \(\lambda\) values fall in a similar range for both metrics, demonstrating the importance of AN power allocation for SEE.


\section{Conclusion} \label{ChapVIConclusion}

This paper studied the secrecy performance of a cooperative SWIPT-NOMA system with AN-aided relaying under practical impairments. In the considered system, the source employs OSTBC transmission, legitimate users adopt RAS, and the EVE applies MRC. The effects of CEEs, FBD, ipSIC, and imperfect AN cancelation are incorporated into the analysis to reflect practical transmission conditions. \textcolor{black}{Analytical SOP expressions and high-SNR secrecy characterizations} are derived and secrecy throughput and SEE are also characterized.
The analytical results are verified by Monte Carlo simulations, showing a close agreement between theory and simulation. The results demonstrated that the proposed scheme benefits from both transmit diversity provided by OSTBC and receive diversity provided by RAS. Increasing the OSTBC order and the number of receive antennas at the legitimate users improves the secrecy performance, whereas increasing the number of EVE antennas degrades it due to the enhanced combining capability of EVE. It is also shown that the AN power allocation factor has a significant effect on the system performance. An appropriate choice of \(\lambda\) improves the secrecy performance by degrading the EVE link while maintaining sufficient power for AF transmission.
Under ideal \textcolor{black}{CSI} conditions, the SOP decreases with increasing transmit SNR, and the asymptotic results accurately characterize the high-SNR behavior. In contrast, under practical impairments, the SOP curves converge to non-zero error floors, which indicates that residual interference and channel uncertainty limit the achievable secrecy performance. In particular, ipSIC causes noticeable degradation for the users performing SIC, while imperfect AN cancelation and outdated/erroneous channel information further reduce the secrecy reliability. The results also showed that proper allocation of AN power can alleviate the performance loss caused by additional EVE antennas and improve both secrecy throughput and SEE.
Overall, the considered OSTBC/RAS-based SWIPT-NOMA system with relay-generated AN provides an effective framework for enhancing physical-layer security in energy-constrained cooperative networks. The obtained results highlight the importance of jointly selecting the OSTBC structure, the number of receive antennas at the legitimate users, the PS ratio, and the AN power allocation factor, while accounting for the combining capability of EVE and practical impairments.
\appendices

\section*{Acknowledgment}
This study is supported by the Scientific and Technological Research Council of Turkey (TUBITAK) under Grant Number 121E467. The authors thank TUBITAK for its support.
\section{Derivation of \(F_{\gamma_l}(x)\)} \label{APP1} 
In this appendix, the detailed derivation of the closed-form expression of \(F_{\gamma_l}(x)\) is presented. From \eqref{Equation9}, the CDF of \(\gamma_l\) is written in terms of the CDF of \(X\) and the PDF of \(Y_l\). Therefore, in the first step, the statistical characterizations of \(X\) and \(Y_l\) are obtained. Then, substituting these expressions into \eqref{Equation9}, the closed-form expression of \(F_{\gamma_l}(x)\) is derived.

\textcolor{black}{Let $X=\|\hat{\mathbf{h}}_{SR}^{\tau}\|^{2}$ denote the equivalent S-to-R channel gain. Since the outdated estimated S-to-R channel coefficients $\hat{h}_{SR,i}^{\tau}$, $i=1,\ldots,N_S$, are assumed to have  independent Nakagami-$m$ envelopes, the corresponding power gains $|\hat{h}_{SR,i}^{\tau}|^{2}$ are independent Gamma random variables. Therefore, $X = \sum_{i=1}^{N_S} |\hat{h}_{SR,i}^{\tau}|^{2}$ is the sum of $N_S$ independent Gamma random variables.}
Consequently, \textcolor{black}{for positive integer $m_1$,} the CDF of \(X\) can be expressed as

\begingroup
\begin{equation}
\label{A1}
F_X(x)
=
1-
\exp\left(-\frac{m_1}{\Omega_1}x\right)
\sum_{n=0}^{m_1N_S-1}
\frac{1}{n!}
\left(\frac{m_1}{\Omega_1}x\right)^n,
\quad x\geq 0.
\end{equation}
\endgroup
For an arbitrary legitimate user, \textcolor{black}{let \(V_l=|\hat{\tilde h}_l^\tau|^2=\max_{1\leq j\leq N_l}|\hat{\tilde h}_{l,j}^{\tau}|^2\) denote the effective relay-to-user channel gain after RAS but before NOMA user ordering. Since the available outdated channel estimates $\hat{\tilde h}_l^\tau$ are assumed to have independent Nakagami-$m$ envelopes, the corresponding power gains $|\hat{\tilde h}_l^\tau|^{2}$ follow Gamma distributions. Hence, \textcolor{black}{for positive integer $m_2$,} the CDF and PDF of $V_l$ can be respectively expressed as}

\begingroup
\small
\begin{equation}
\label{A2}
F_{V_l}(x)
=
\left[
1-
\exp\left(-\frac{m_2}{\Omega_2}x\right)
\sum_{r=0}^{m_2-1}
\frac{1}{r!}
\left(\frac{m_2}{\Omega_2}x\right)^r
\right]^{N_l},
\quad x\geq 0,
\end{equation}
\endgroup
\begingroup
\small
\begin{equation}
\label{A3}
\begin{aligned}
f_{V_l}(x)
&=
N_l
\left(\frac{m_2}{\Omega_2}\right)^{m_2}
\frac{1}{\Gamma(m_2)}
\sum_{p=0}^{N_l-1}
\sum_{q=0}^{p(m_2-1)}
\binom{N_l-1}{p}
(-1)^p
\\
&\quad \times
\mu_q(p,m_2) x^{q+m_2-1}
\exp\left(-\frac{m_2(p+1)}{\Omega_2}x\right),
\quad x\geq 0,
\end{aligned}
\end{equation}
\endgroup
where $\mu(\cdot,\cdot)$ are multinomial coefficients. \textcolor{black}{Since the NOMA users are ordered according to their available effective channel gains after the RAS operation, $Y_l=|\hat h_l^\tau|^2$ corresponds to the $l$-th ordered statistic of $V_l$. Therefore, using the theory of ordered statistics, the PDF of $Y_l$ can be expressed as}

\begingroup
\begin{equation}
\label{A4}
\begin{aligned}
f_{Y_l}(x)
&=
Q_l
\sum_{k=0}^{L-l}
(-1)^k
\binom{L-l}{k}
f_{V_l}(x)
\left(F_{V_l}(x)\right)^{l+k-1}
\\
&=
\frac{Q_l N_l\left(\dfrac{m_2}{\Omega_2}\right)^{m_2}}{\Gamma(m_2)}
\sum_{k,p,q,r,s} 
C^{L-l}_{k}
C^{N_l-1}_{p}
C^{N_l(l+k-1)}_{r}\\
& \times(-1)^{k+p+r}
\mu_q(p,m_2)\mu_s(r,m_2)
 x^{\phi_1-1}\\
& \times
\exp\left(
-\frac{m_2(p+r+1)}{\Omega_2}x
\right),
\end{aligned}
\end{equation}
\endgroup
where $\sum_{k,p,q,r,s} = \sum_{k=0}^{L-l}
\sum_{p=0}^{N_l-1}
\sum_{q=0}^{p(m_2-1)}
\sum_{r=0}^{N_l(l+k-1)}$ $\sum_{s=0}^{r(m_2-1)}$, \(Q_l=\frac{L!}{(l-1)!(L-l)!}\), and \(C_x^y \triangleq \frac{x!}{y!(x-y)!}\) represent the binomial coefficient. By substituting \eqref{A1} and \eqref{A4} into \eqref{Equation9}, the CDF of \(\gamma_l\) can be expressed as

\begingroup
\begin{equation}
\label{A5}
\begin{aligned}
F_{\gamma_l}(x)
&=
1-
\frac{
Q_l N_l
\left(\dfrac{m_2}{\Omega_2}\right)^{m_2}
}{
\Gamma(m_2)
}
\sum_{k,p,q,r,s,n} C^{L-l}_{k}
C^{N_l-1}_{p}
C^{N_l(l+k-1)}_{r} \\
& \times
\frac{(-1)^{k+p+r}}{n!}
\left(\frac{m_1}{\Omega_1}\right)^n
\mu_q(p,m_2)\mu_s(r,m_2)
\\
&\times
\int_{\frac{N_S R_c\kappa_3x}{A_l(x)}}^{\infty}
\left(\chi_l(x,y)\right)^n
\exp\left(-\frac{m_1}{\Omega_1}\chi_l(x,y)\right)
\\
& \quad \times y^{\phi_1-1}
\exp\left(-\frac{m_2(p+r+1)}{\Omega_2}y\right)
dy,
\end{aligned}
\end{equation}
\endgroup
where $\chi_l(x,y)=\frac{N_S R_c x\left[By+(\bar{\gamma}\kappa_4+\kappa_6)\right]}{\bar{\gamma}\left[A_l(x)y-N_S R_c\kappa_3x\right]}$ with $A_l(x)=a_l-\kappa_1x$ and $B=\bar{\gamma}\kappa_2+\kappa_5$. Then, the change of variable $u=A_l(x)y-N_S R_c\kappa_3x$
is employed. Under this transformation, the term $By+(\bar{\gamma}\kappa_4+\kappa_6)$ leads to the compact expression $D_l(x)=BN_S R_c\kappa_3x+(\bar{\gamma}\kappa_4+\kappa_6)A_l(x)$, which is used in \eqref{Equation10}. Finally, using the binomial expansion and the integral identity given in \cite[eq.~(3.471.9)]{Grad2000}, the closed-form expression in \eqref{Equation10} is obtained.

\section{Derivation of the PDF of \(\gamma_{E,l}\)} \label{APP2} 
In this appendix, the detailed derivation of \(f_{\gamma_{E,l}}(x)\) is presented. By using the instantaneous SINR expression at EVE given in \eqref{Equation6}, the CDF \(F_{\gamma_{E,l}}(x)\) is first derived. Then, the PDF \(f_{\gamma_{E,l}}(x)\) is obtained by differentiating \(F_{\gamma_{E,l}}(x)\) with respect to \(x\). Since \(Z=\|\mathbf{h}_{E}\|^2=\sum_{i=1}^{N_E}|h_{E,i}|^2\) is the sum of \(N_E\) independent and identically distributed (i.i.d.) Gamma random variables, it also follows a Gamma distribution. Therefore, the PDF of \(Z\) can be expressed as

\begingroup
\begin{equation}
\label{B1}
f_Z(z)
=
\left(\frac{m_E}{\Omega_E}\right)^{m_E N_E}
\frac{z^{m_E N_E-1}}{\Gamma(m_E N_E)}
\exp\left(-\frac{m_E}{\Omega_E}z\right),
\quad z\geq 0.
\end{equation}
\endgroup

Then, by using \eqref{Equation6}, the CDF of \(\gamma_{E,l}\) can be expressed as

\begingroup
\begin{equation}
\label{EquationB2}
\begin{aligned}
F_{\gamma_{E,l}}(x)
&=
\Pr\left(\gamma_{E,l}\leq x\right)
\\
&=
\Pr\left(
X \leq \chi_{E,l}(x,z)
,
\;
Z>0
\right) \\ &=
\int_{0}^{\infty}
F_X\left(
\chi_{E,l}(x,z)
\right)
f_Z(z)dz,
\end{aligned}
\end{equation}
\endgroup
which is valid for \(x<\frac{a_l}{\kappa_{1e}}\).  Otherwise, it follows that \(F_{\gamma_{E,l}}(x)=1\). Here, $\chi_{E,l}(x,y)=\frac{N_S R_c x\left(B_e Z+\kappa_{4e}\right)}{\bar{\gamma}_E A_{e,l}(x)Z}$ with $A_{e,l}(x)=a_l-\kappa_{1e}x$ and $B_e=\bar{\gamma}_E\kappa_{2e}+\kappa_{3e}$.
Substituting the CDF of \(X\) given in \eqref{A1} and the PDF of \(Z\) given in \eqref{B1} into \eqref{EquationB2}, \(F_{\gamma_{E,l}}(x)\) can be written as

\begingroup
\begin{equation}
\label{EquationB3}
\begin{aligned}
F_{\gamma_{E,l}}(x)
&=
1-
\frac{
\left(\dfrac{m_E}{\Omega_E}\right)^{m_E N_E}
}{
\Gamma(m_E N_E)
}
\sum_{t_1=0}^{m_1N_S-1}
\frac{1}{t_1!}
\left(
\frac{m_1N_SRx}
{\Omega_1\bar{\gamma}_E A_{e,l}(x)}
\right)^{t_1}
\\
& \times
\int_{0}^{\infty}
\exp\left(
-\frac{
m_1 \chi_{E,l}(x,y) }{\Omega_1 z}
\right)
\left(B_ez+\kappa_{4e}\right)^{t_1}
\\
& \times z^{m_E N_E-t_1-1}
\exp\left(-\frac{m_E}{\Omega_E}z\right)
dz.
\end{aligned}
\end{equation}
\endgroup

Then, by applying the binomial expansion to the term \(\left(B_ez+\kappa_{4e}\right)^{t_1}\), the integral in \eqref{EquationB3} can be rewritten as a finite summation of standard exponential-type integrals. Subsequently, using the identity in \cite[eq.~(3.471.9)]{Grad2000}, the resulting integrals are evaluated in closed form. Hence, after some algebraic manipulations, the CDF of \(\gamma_{E,l}\) is obtained as

\begingroup
\begin{equation}
\label{B4}
\begin{aligned}
F_{\gamma_{E,l}}(x)
&=
1-
\frac{
\left(\dfrac{m_E}{\Omega_E}\right)^{m_E N_E}
}{
\Gamma(m_E N_E)
}
\sum_{t_1,t_2}
C^{t_1}_{t_2}
\frac{2\kappa_{4e}^{t_2}}{t_1!}
\left(\frac{m_1}{\Omega_1}\right)^{t_1}
B_e^{t_1-t_2}\\
& \times
\left(\frac{N_SR}{\bar{\gamma}_E}\right)^{t_1}
\left(
\frac{
N_SRm_1\Omega_E\kappa_{4e}
}{
\bar{\gamma}_E\Omega_1m_E
}
\right)^{\frac{\psi_E}{2}} \left(
\frac{x}{A_{e,l}(x)}
\right)^{\frac{\psi_E}{2}+t_1} \\ &\times
\exp\left(
-\frac{
N_SRm_1B_ex
}{
\bar{\gamma}_E\Omega_1A_{e,l}(x)
}
\right)
K_{\psi_E}\!\left(\Delta_e(x)\right),
\end{aligned}
\end{equation}
\endgroup
where $\sum_{t_1, t_2}=\sum_{t_1=0}^{m_1N_S-1}\sum_{t_2=0}^{t_1}$, $\psi_E=m_EN_E-t_2$, $\Delta_e(x)=2\sqrt{\frac{N_SRm_1m_E\kappa_{4e}}{\bar{\gamma}_E\Omega_1\Omega_E}\frac{x}{A_{e,l}(x)}}$, and \(K_v(\cdot)\) is the modified Bessel function of the second order \(v\).
Then, by taking the first-order derivative of \(F_{\gamma_{E,l}}(x)\) with respect to \(x\) and performing some algebraic manipulations, the PDF \(f_{\gamma_{E,l}}(x)\) is obtained as given in \eqref{Equation11}.

\section{\textcolor{black}{Derivation of the Asymptotic $P_{S,l}^{\infty}$}}
\label{APP3}

\textcolor{black}{
In this appendix, the coefficients of the asymptotic CDF in
\eqref{Equation15} are first provided, followed by a tractable
Gaussian--Chebyshev representation of the asymptotic SOP in
\eqref{EquationXXasimp}. The coefficients corresponding to the three
cases in \eqref{Equation15} are given as follows. For $M<D_l$,
}

\begingroup
\small
\begin{equation}
\label{C1}
\begin{aligned}
\Upsilon_{1,l}
&=
\frac{
Q_lN_l
\left(\dfrac{m_2}{\Omega_2}\right)^{m_2}
}{
\Gamma(m_2)
}
\frac{
\left(\dfrac{m_1}{\Omega_1}\right)^M
}{
\Gamma(M+1)
}
\sum_{n=0}^{M}
C_n^M \nu_l^n
\sum_{k,p,q,r,s}
C_k^{L-l}
C_p^{N_l-1}\\
&\quad\times
C_r^{N_l(l+k-1)}
(-1)^{k+p+r}
\mu_q(p,m_2)\mu_s(r,m_2)
\\
&\quad\times
\mathcal{H}
\left(
\phi_1-n,
\frac{m_2(p+r+1)}{\Omega_2}
\right),
\end{aligned}
\end{equation}
\endgroup
\textcolor{black}{
where $\sum_{k,p,q,r,s}$ follows the same summation ranges defined
in \eqref{A4}, $\phi_1=m_2+q+s$, and $\mathcal H(a,b)= \begin{cases} \Gamma(a)b^{-a}, & a>0,\\[1mm] \dfrac{(-1)^j b^j}{j!}\,[\operatorname{dg}(j+1)-\ln b], & a=-j,\;j=0,1,\ldots .\end{cases}$ Here, $\operatorname{dg}(x)=\Gamma'(x)/\Gamma(x)$ denotes the
digamma function given in \cite[eq.~(6.3.1)]{Abram1972}.
For $M>D_l$, the corresponding coefficient is
$\Upsilon_{2,l}
=
C_l^L
\frac{
\left(\dfrac{m_1}{\Omega_1}\right)^{D_l}
\left(\dfrac{m_2}{\Omega_2}\right)^{D_l}
}{
\left[\Gamma(m_2+1)\right]^{lN_l}
}
\nu_l^{D_l}
\frac{\Gamma(M-D_l)}{\Gamma(M)}$.
Finally, for $M=D_l=d$, the coefficient associated with the
logarithmic term in \eqref{Equation15} is given by
$\Upsilon_{3,l}
=
C_l^L
\frac{
d\nu_l^d
\left(\dfrac{m_1}{\Omega_1}\right)^d
\left(\dfrac{m_2}{\Omega_2}\right)^d
}{
\Gamma(d+1)
\left[\Gamma(m_2+1)\right]^{lN_l}
}$.
}
\textcolor{black}{
Using these coefficients, a numerically tractable representation of
the asymptotic SOP in \eqref{EquationXXasimp} can be obtained.
Since $F_{\gamma_l}^{\infty}(x)=1$ for
$x\geq a_l/\kappa_1$, the effective integration limit is
$\varrho^\star
=
\min\left\{
\frac{1}{\psi_l}
\left(
\frac{a_l}{\kappa_1}+1
\right)-1,\,
\frac{a_l}{\kappa_{1e}}
\right\}$. Then, using the transformation
$\gamma=\frac{\varrho^\star}{2}(u+1)$ and the
Gaussian--Chebyshev quadrature formula in
\cite[eq.~(25.4.38)]{Abram1972}, the asymptotic SOP in
\eqref{EquationXXasimp} is approximated as
}

\begingroup
\begin{equation}
\label{C6}
\begin{aligned}
P_{S,l}^{\infty}
&\approx
\frac{\varrho^\star\Upsilon_l}{2}
\sum_{i=1}^{I}
w_i\sqrt{1-u_i^2}
\\
&\quad\times
\left[
\tau_l\!\left(\Phi_l(\gamma_i)\right)
\right]^{d_l^\star}
\left[
\ln\left(
\frac{1}{
\tau_l\!\left(\Phi_l(\gamma_i)\right)}
\right)
\right]^{\delta_l}
f_{\gamma_{E,l}}(\gamma_i),
\end{aligned}
\end{equation}
\endgroup

\textcolor{black}{
where
$u_i=\cos\left(\frac{(2i-1)\pi}{2I}\right)$,
$\gamma_i=\frac{\varrho^\star}{2}(u_i+1)$, and
$w_i=\frac{\pi}{I}$.
}
\bibliographystyle{IEEEtran}
\bibliography{ref}

\end{document}